\documentclass[aps, prd, twocolumn, superscriptaddress, nofootinbib, floatfix, noshowpacs]{revtex4-2}

\usepackage{amsmath,amssymb}
\usepackage{color,url}
\usepackage{listings}
\usepackage{slashed}
\usepackage[pdftex]{graphicx}
\usepackage{epstopdf}
\usepackage{epsfig}
\usepackage{grffile}
\usepackage{relsize}
\usepackage{float}
\graphicspath{{./img/}}
\usepackage{soul}

\usepackage{CJKutf8}

\usepackage[mathscr,scaled=1.15]{urwchancal}
\DeclareFontFamily{OT1}{pzc}{}
\DeclareFontShape{OT1}{pzc}{m}{it}%
{<-> s * [1.15] pzcmi7t}{}
\DeclareMathAlphabet{\mathpzc}{OT1}{pzc}{m}{it}

\definecolor{purple}{rgb}{0.5,0,0.5}
\definecolor{blue}{rgb}{0.0,0,0.9}
\definecolor{prdblue}{rgb}{0.133,0.118,0.498}
\usepackage[colorlinks=true, pdfstartview=FitV, linkcolor=prdblue, citecolor= prdblue, urlcolor=prdblue]{hyperref}

\RequirePackage{comment}

\begin{document}
\begin{CJK*}{UTF8}{gbsn}

\title{$\,$\\[-6ex]\hspace*{\fill}{\normalsize{\sf\emph{Preprint no}.\
NJU-INP 107-25}}\\[1ex]
Light-Front Transverse Nucleon Charge and Magnetisation Densities}

\author{Z.-N. Xu (徐珍妮)%
    $^{\href{https://orcid.org/0000-0002-9104-9680}{\textcolor[rgb]{0.00,1.00,0.00}{\sf ID}}}$}
\affiliation{Dpto.\ Ciencias Integradas, Centro de Estudios Avanzados en Fis., Mat.\ y Comp., \\
\hspace*{0.5em}Fac.~Ciencias Experimentales, \href{https://ror.org/03a1kt624}{Universidad de Huelva}, E-21071 Huelva, Spain}

\author{Z.-Q. Yao (姚照千)%
       $\,^{\href{https://orcid.org/0000-0002-9621-6994}{\textcolor[rgb]{0.00,1.00,0.00}{\sf ID}}}$}
\email{z.yao@hzdr.de}
\affiliation{\href{https://ror.org/01zy2cs03}{Helmholtz-Zentrum Dresden-Rossendorf}, Bautzner Landstra{\ss}e 400, D-01328 Dresden, Germany}

\author{
    P.\ Cheng (程鹏)%
    $\,^{\href{https://orcid.org/0000-0002-6410-9465}{\textcolor[rgb]{0.00,1.00,0.00}{\sf ID}}}$}
\affiliation{Department of Physics, \href{https://ror.org/05fsfvw79}{Anhui Normal University}, Wuhu, Anhui 24100, China}

\author{C. D. Roberts%
       $^{\href{https://orcid.org/0000-0002-2937-1361}{\textcolor[rgb]{0.00,1.00,0.00}{\sf ID}}}$}
\email[]{cdroberts@nju.edu.cn}
\affiliation{School of Physics, \href{https://ror.org/01rxvg760}{Nanjing University}, Nanjing, Jiangsu 210093, China}
\affiliation{Institute for Nonperturbative Physics, \href{https://ror.org/01rxvg760}{Nanjing University}, Nanjing, Jiangsu 210093, China}

\author{J. Rodr\'iguez-Quintero%
       $^{\href{https://orcid.org/0000-0002-1651-5717}{\textcolor[rgb]{0.00,1.00,0.00}{\sf ID}}}$}
\email{jose.rodriguez@dfaie.uhu.es}
\affiliation{Dpto.\ Ciencias Integradas, Centro de Estudios Avanzados en Fis., Mat.\ y Comp., \\
\hspace*{0.5em}Fac.~Ciencias Experimentales, \href{https://ror.org/03a1kt624}{Universidad de Huelva}, E-21071 Huelva, Spain}

\author{J.\ Segovia%
    $^{\href{https://orcid.org/0000-0001-5838-7103}{\textcolor[rgb]{0.00,1.00,0.00}{\sf ID}}}$}
\affiliation{Dpto.\ Sistemas F\'{\i}sicos, Qu\'{\i}micos y Naturales, \href{https://ror.org/02z749649}{Univ.\ Pablo de Olavide}, E-41013 Sevilla, Spain}

\date{2026 February 19}  

\begin{abstract}
Nucleon elastic electromagnetic form factors obtained using both the three-body and quark + fully-interacting-diquark pictures of nucleon structure are employed to calculate an array of light-front transverse densities for the proton and neutron and their dressed valence-quark constituents, \emph{viz}.\ flavour separations of the proton and neutron results.  These two complementary descriptions of nucleon structure deliver mutually compatible predictions, which match expectations based on modern parametrisations of available data, where such are available.  Amongst other things, it is found that transverse-plane valence $u$- and $d$-quark Dirac radii are practically indistinguishable; but regarding kindred Pauli radii, the $d$ quark value is roughly 10\% greater than that of the $u$-quark.  Moreover, magnetically, the valence $d$ quark is far more active than the valence $u$ quark, probably because it has much greater orbital angular momentum.  Both pictures of nucleon structure agree in predicting that, in a polarised nucleon, the transverse-plane charge densities are no longer rotationally invariant.  Instead, for a $+\hat x$ polarised nucleon, positive charge is displaced in the $+\hat y$ direction, with the opposite effect for negative charge.
\end{abstract}
\maketitle
\end{CJK*}

\section{Introduction}
Drawing charts of nuclear and nucleon structure has long been a principal goal of high energy nuclear and particle physics.  Focusing on nucleons, the Poincar\'e-invariant electromagnetic form factors contain much objective information.  Recognising this, a great deal of electron + nucleon scattering data has been collected and analysed since the pioneering experiments described in Ref.\,\cite{Hofstadter:1956qs}.
That progress is canvassed in, \emph{e.g}., Refs.\,\cite{Perdrisat:2006hj, Punjabi:2015bba, Gao:2021sml, Cui:2022fyr}.

Understanding and relating modern data to a sound picture of nucleon structure requires robust theory.  Since the data now extend to large values of spacelike squared momentum transfer, \emph{viz}.\ $0<Q^2 \lesssim 15\,$GeV$^2$ \cite{Gilfoyle:2018xsa, Wojtsekhowski:2020tlo}, a coherent description requires a Poincar\'e-covariant approach because one is delving deep into a target with highly relativistic constituents.

One such scheme is provided by lattice-regularised quantum chromodynamics (lattice-QCD or lattice Schwinger function methods).
Material progress has been made in developing lattice-QCD for form factor calculations since the turn of this century; see, \emph{e.g}., Refs.\,\cite{Yamazaki:2009zq, Bhattacharya:2013ehc, QCDSF:2017ssq, Alexandrou:2018sjm, Alexandrou:2023qor, Syritsyn:2025fiu}.
Notwithstanding that, reaching the large-$Q^2$ domain now accessible to experiments remains difficult using lattice-QCD, owing to the need for very fine lattices and large volumes, and issues with signal-to-noise ratio and excited state contamination \cite{Syritsyn:2025fiu}.

Hamiltonian-based light-front approaches and relativistic and/or holographic quark models have also been used \cite{deMelo:2008rj, Giannini:2015zia, Sufian:2016hwn, Mondal:2019jdg, Ahmady:2021qed, Xu:2021wwj} with some success.
The issue in these cases, however, is forging a link to QCD because the degrees of freedom used to formulate the models do not have a traceable connection to those in the QCD action.

A viable alternative to the aforementioned approaches is provided by continuum Schwinger function methods (CSMs) \cite{Eichmann:2016yit, Burkert:2019bhp, Binosi:2022djx, Ding:2022ows, Ferreira:2023fva, Achenbach:2025kfx}.
In this framework, one typically works with QCD's Euler-Lagrange equations of motion -- the Dyson-Schwinger equations (DSEs) -- and develops symmetry preserving truncation schemes that enable a systematic connection to be made between each element of a given calculation and fields and/or terms in the QCD action.
Regarding nucleon elastic electromagnetic form factors, two complementary paths are currently in use.
Namely, a direct approach based on a $3$-body Faddeev equation for the nucleon wave function \cite{Eichmann:2011vu, Yao:2024uej}
and a quark + fully-interacting diquark -- $q(qq)$ -- picture of nucleon structure \cite{Barabanov:2020jvn, Cheng:2025yij}.
Both schemes are Poincar\'e-covariant and can deliver form factors that cover and reach beyond the range of existing experiments.

One particular use of nucleon electromagnetic form factors is to produce charts of the distribution of charge and magnetisation within these states.
Drawing analogies with nonrelativistic systems, three-dimensional Fourier transforms of the Poincar\'e-invariant Sachs electric and magnetic form factors \cite{Sachs:1962zzc}, \emph{viz}.\ $G_{E,M}^N(Q^2)$, when interpreted in the Breit frame, were long imagined to describe nucleon configuration space charge and magnetisation densities.
However, this interpretation overlooks the fact that Poincar\'e-covariant wave functions do not have a probability interpretation; hence, cannot be used to define densities.

\begin{figure}[!t]
	\centering
	\includegraphics[width=8cm]{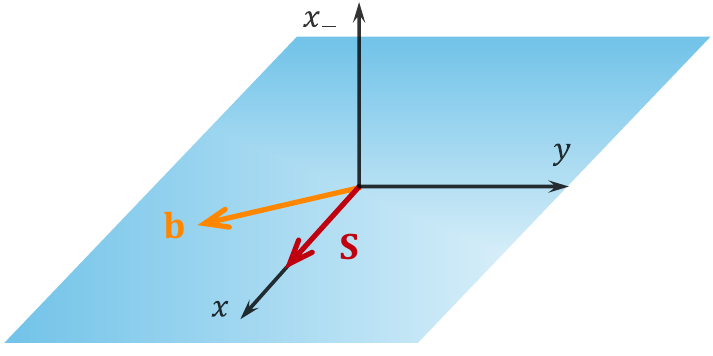}
	\caption{%
With the nucleon's direction of motion defining the longitudinal light-front vector, the light-front transverse plane can be characterised by the usual two cartesian vectors, $(\hat x,\hat y)$, as marked by the blue surface in the image.
The transverse position vector is indicated by $\vec{b}$ -- orange; and the transverse spin vector by $\vec{S}$ -- red.
	\label{IMF}}
	\end{figure}

A contemporary alternative \cite{Miller:2010nz} considers projections of the Poincar\'e-invariant form factors onto the light-front via two-dimensional Fourier transforms with $Q^2$ interpreted in a Drell-Yan-West frame \cite{Drell:1969km, West:1970av}.
Such functions describe true densities in a plane perpendicular to the light-front vector used in defining the nucleon state; see Fig.\,\ref{IMF}.  Interpreted in the infinite momentum frame, this is the nucleon's direction of motion.  No information is available about light-front-longitudinal charge and magnetisation distributions.

Herein, we calculate and compare transverse charge densities obtained using both the $3$-body \cite{Yao:2024uej} and $q(qq)$ \cite{Cheng:2025yij} approaches to nucleon structure.
Sections~\ref{S:Schwinger} and \ref{NCurrent} provide a detailed discussion of the contemporary $3$-body framework for calculation of nucleon form factors.  This scheme was employed in Ref.\,\cite{Yao:2024uej}, but little in the way of explanation was provided.  
%
%
An exhaustive background to the $q(qq)$ approach is provided in Ref.\,\cite{Barabanov:2020jvn} and a complete exposition of the modern implementation was provided in Ref.\,\cite{Cheng:2025yij}, so little recapitulation is necessary.
Transverse charge and magnetisation densities, and the kindred charge density associated with a transversely polarised nucleon, are defined, and the CSM predictions reported and discussed in Sec.\,\ref{S:density}.
Section~\ref{S:S} provides a summary and perspective.

\begin{figure}[t]
\centerline{%
\includegraphics[clip, width=0.44\textwidth]{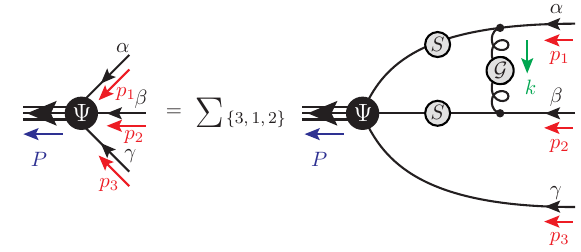}}
\caption{\label{FigFaddeev}
Faddeev equation in RL truncation.
Filled circle: Faddeev amplitude, $\Psi$, the matrix-valued solution, which involves 128 independent scalar functions.
Spring: dressed-gluon interaction that mediates quark+quark scattering; see Eqs.\,\eqref{EqRLInteraction}, \eqref{defcalG}.
Solid line: dressed-quark propagator, $S$, calculated from the rainbow gap equation.
Lines not adorned with a shaded circle are amputated.
Isospin symmetry is assumed.
The sum runs over each of the cases involving quark ``$i=1,2,3$'' as a spectator to the exchange interaction.}
\end{figure}

\section{\label{S:Schwinger}Faddeev Equation}
\subsection{Basic elements}
Herein, we work at leading-order in a systematically-improvable, symmet\-ry-preserving truncation of all quantum field (Dyson-Schwinger) equations which appear in the matrix element that defines the photon + nucleon interaction; namely, the rainbow-ladder (RL) truncation \cite{Munczek:1994zz, Bender:1996bb}.
It is worth mentioning here that, after roughly thirty years of use, RL truncation is known to be reliable for pion, kaon, and nucleon observables:
(\emph{a}) practically, owing to numerous successful applications \cite{Roberts:2021nhw, Ding:2022ows, Raya:2024ejx};
and (\emph{b}) because improvement schemes exist and have been tested, showing that the cumulative effect of improvements to RL truncation in these channels can be absorbed into a modest modification of the quark + quark scattering kernel \cite{Fischer:2009jm, Chang:2009zb, Chang:2013pq, Binosi:2014aea, Williams:2015cvx, Binosi:2016rxz, Binosi:2016wcx, Xu:2022kng}.
Importantly, where reasonable comparisons are possible, modern CSM predictions and contemporary lattice-QCD results are mutually consistent -- see, \emph{e.g}., Refs.\,\cite{Roberts:2021nhw, Binosi:2022djx, Ding:2022ows, Ferreira:2023fva, Raya:2024ejx, Chen:2021guo, Chang:2021utv, Lu:2023yna, Chen:2023zhh, Yu:2024qsd, Alexandrou:2024zvn}.

In RL truncation, the nucleon bound state amplitude can be obtained by solving the Faddeev equation sketched in Fig.\,\ref{FigFaddeev}.
Two elements are required to complete the definition of the Faddeev equation:
(\emph{i}) the quark + quark interaction, $\tilde{\mathpzc G}$;
and (\emph{ii}) the dressed quark Schwinger function (propagator), $S$.

In RL truncation, the scattering kernel is obtained by writing \cite{Maris:1997tm}:
{\allowdisplaybreaks
%
\begin{equation}
\label{EqRLInteraction}
\mathscr{G}_{CD}^{AB}(k)  =
\tilde{\mathpzc G}(y)
 [i\gamma_\mu\frac{\lambda^{a}}{2} ]_{CB} [i\gamma_\nu\frac{\lambda^{a}}{2} ]_{DA}
 T_{\mu\nu}^k\,,
\end{equation}
$k^2 T_{\mu\nu}^k = k^2 \delta_{\mu\nu} - k_\mu k_\nu$,  $y=k^2$,
$A, B, C, D$ represent colour and spinor matrix indices.
The $T_{\mu\nu}$ specifies Landau gauge, employed because it is a fixed point of the renormalisation group and also that gauge for which corrections to RL truncation are least significant \cite{Bashir:2009fv}.
}

A realistic form of ${\mathpzc G}_{\mu\nu}(y)$ is described in Refs.\,\cite{Qin:2011dd, Binosi:2014aea}.  It is specified by writing
\begin{align}
\label{defcalG}
 \tilde{\mathpzc G}(y) & =
 \frac{8\pi^2}{\omega^4} D e^{-y/\omega^2} + \frac{8\pi^2 \gamma_m \mathcal{F}(y)}{\ln\big[ \tau+(1+y/\Lambda_{\rm QCD}^2)^2 \big]}\,,
\end{align}
where $\gamma_m=12/25$, $\Lambda_{\rm QCD} = 0.234\,$GeV, $\tau={\rm e}^2-1$, and ${\cal F}(y) = \{1 - \exp(-y/\Lambda_{\mathpzc I}^2)\}/y$, $\Lambda_{\mathpzc I}=1\,$GeV.
%
%
Asymptotic freedom is manifest in the ultraviolet behaviour of $\tilde{\mathpzc G}(y)$ and confinement is expressed throughout via the violation of reflection positivity by coloured Schwinger functions \cite[Sec.\,5]{Ding:2022ows}.

In formulating and solving the relevant equations, we use a mass-independent (chiral-limit) momentum-subtraction renormalisation scheme \cite{Chang:2008ec}, with renormalisation scale $\zeta=\zeta_{19} = 19\,$GeV, \emph{viz}.\ a far ultraviolet value so as to avoid truncation ambiguity and ensure that the the quark wave function renormalisation constant is essentially unity.

Numerous applications have revealed \cite{Ding:2022ows} that interactions in the class containing Eqs.\,\eqref{EqRLInteraction}, \eqref{defcalG} can serve to unify the properties of many systems.
Contemporary studies employ $\omega = 0.8\,$GeV \cite{Xu:2022kng}.
Then, with $\omega D = 0.8\,{\rm GeV}^3$ and renormalisation point invariant quark current mass $\hat m_u = \hat m_d = 6.04\,$MeV, which corresponds to a one-loop evolved mass at $\zeta=2\,$GeV of $4.19\,$MeV, the following predictions are obtained: pion mass $m_\pi = 0.14\,$GeV; nucleon mass $m_N=0.94\,$GeV; and pion leptonic decay constant $f_\pi=0.094\,$GeV.  These values are a good match with experiment \cite{ParticleDataGroup:2024cfk}.
%
%
When the product $\omega D$ is held constant, physical observables remain practically unchanged under $\omega \to (1\pm 0.2)\omega$ \cite{Qin:2020rad}.

It is worth highlighting that all results obtained subsequently are parameter-free predictions.  The interaction involves one parameter and there is a single quark current-mass.  At this point, both quantities are fixed.

Here, it is also worth providing additional information about the interaction in Eq.\,\eqref{defcalG}.
For instance, following Ref.\,\cite{Qin:2011dd}, one may draw a connection between $\tilde{\mathpzc G}$
and QCD's process-independent (PI) effective charge \cite{Cui:2019dwv, Deur:2023dzc, Brodsky:2024zev}.
That PI charge is characterised by an infrared coupling value $\hat\alpha(0)/\pi = 0.97(4)$ and a gluon mass-scale $\hat m_0 = 0.43(1)\,$GeV, both determined together via a combined continuum and lattice analysis of QCD's gauge sector \cite{Cui:2019dwv}.
The following values are those of analogous quantities inferred from Eq.\,(2): 
\begin{equation}
\label{DiscussInteraction}
\alpha_{\mathpzc G}(0)/\pi = 1.45\,,\quad m_{\mathpzc G} = 0.54\,{\rm GeV}\,.
\end{equation}
They agree fairly well with the QCD values.

Element (\emph{ii}) in Fig.\,\eqref{FigFaddeev} is the dressed quark propagator, which has the form
\begin{equation}
S(k;\zeta) = Z(k^2;\zeta)/[i \gamma\cdot k + M(k^2)]\,,
\end{equation}
where $Z$ is the quark wave function renormalisation function and $M$ is the renormalisation group invariant quark mass function.
Herein, as required by symmetry considerations, it is obtained by solving the RL truncation quark gap (Dyson) equation.
In accomplishing that task, we follow the procedures explained in Refs.\,\cite{Maris:1997tm, Maris:2005tt, Krassnigg:2009gd}.

\subsection{Faddeev Amplitude}
The solution of the three valence body equation sketched in Fig.\,\ref{FigFaddeev} has the following form:
\begin{equation}
\begin{aligned}\label{faveq}
\Psi_{A B}^{C D}(p, q, P)&= \sum_{i=1}^3\Psi^{(i)}_{A B C D}(p, q, P)\,,
\end{aligned}
\end{equation}
where
{\allowdisplaybreaks
\begin{subequations}
\begin{align}
		\Psi^{(1)}_{A B C D}(p, q, P)&= \int_{dk}\mathcal{G}^{C^\prime B^\prime}_{B\ C}(k) {S}_{B^{\prime} B^{\prime \prime}}(k_{2}) {S}_{C^{\prime} C^{\prime \prime}}(\tilde{k}_{3}) \nonumber \\
		&\quad \times\Psi_{A B^{\prime \prime}}^{C^{\prime \prime} D}(p^{(1)}, q^{(1)}, P)\,,\\
		\Psi^{(2)}_{A B C D}(p, q, P)& = \int_{dk}\mathcal{G}^{A^\prime C^\prime}_{C\ A}(k){S}_{C^{\prime} C^{\prime \prime}}(k_{3}) {S}_{A^{\prime} A^{\prime \prime}}(\tilde{k}_{1}) \nonumber\\
		&\quad \times\Psi_{A^{\prime \prime} B}^{C^{\prime \prime} D}(p^{(2)}, q^{(2)}, P)\,, \\
		\Psi^{(3)}_{A B C D}(p, q, P) & =\int_{dk}\mathcal{G}^{B^\prime A^\prime}_{A\ B}(k){S}_{A^{\prime} A^{\prime \prime}}(k_{1}) {S}_{B^{\prime} B^{\prime \prime}}(\tilde{k}_{2})\nonumber \\
& \quad \times\Psi_{A^{\prime \prime} B^{\prime \prime}}^{C D}(p^{(3)}, q^{(3)}, P)\,.
\label{faveq3c}
	\end{align}
	\label{faveq3}
\end{subequations}
\hspace*{-0.5\parindent}In Eq.\,\eqref{faveq3},
$\int_{dk}$ represents a translationally invariant regularisation of the four-dimensional momentum-space integral and we have extended the range of the indices $A, B, C, D$ to include spinor ($\alpha, \beta, \gamma, \delta$), colour ($r, s, t, u$), and flavour ($a, b, c, d)$, with $D$ labelling these quantities for the nucleon itself.
}

Three independent particle momenta are involved in Eq.\,\eqref{faveq3}: $p_{1}$, $p_{2}$, $p_{3}$.
Momentum conservation requires $p_1+p_2+p_3=P$, the total nucleon momentum.
The other two (relative) momenta can be chosen in any way one desires; however, if one wishes to capitalise on permutation symmetry, then the following choices are best \cite{Eichmann:2011vu}:
\begin{align}
p&=(1-\xi) p_{3}-\xi\left(p_{1}+p_{2}\right);& p_{1}&=-q-\frac{p}{2}+\frac{1-\xi}{2} P, \nonumber \\
q&=\frac{p_{2}-p_{1}}{2};& p_{2}&=q-\frac{p}{2}+\frac{1-\xi}{2} P, \nonumber \\
P&=p_{1}+p_{2}+p_{3};& p_{3}&=p+\xi P\,, \label{Involvesxi}
\end{align}
with $\xi=1 / 3$ being a momentum partitioning parameter.
(We have verified that the results are independent of this choice, but the chosen value speeds the calculations.)
The internal quark propagators depend on the quark momenta $k_{i}=p_{i}-k$ and $\tilde{k}_{i}=p_{i}+k$, with $k$ being the exchanged momentum.
For each of the three terms in Eq.\,\eqref{faveq3}, the internal relative momenta are:
\begin{equation}
\begin{aligned}
p^{(1)} & = p+k\,, & p^{(2)}&=p-k\,, &  p^{(3)}&=p\,, \\
q^{(1)}&=q-k / 2\,, &  q^{(2)}&=q-k / 2\,, &  q^{(3)}&=q+k\,.
\end{aligned}
\end{equation}


\begin{figure*}[t]
\centerline{%
\includegraphics[clip, width=0.86\textwidth]{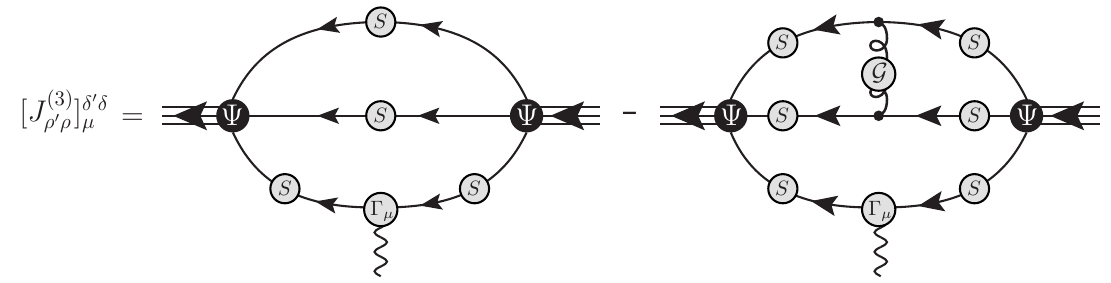}}
\caption{\label{FigCurrent}
The nucleon has three valence quarks, so the complete RL nucleon electromagnetic current has three terms: $J_\mu(Q) = \sum_{a=1,2,3} J_\mu^a(Q)$.
Using symmetries, one can readily obtain the $a=1,2$ components once the $a=3$ component is known \cite[Appendix~B]{Eichmann:2011pv}.
%
$\delta$, $\delta^\prime$ are spinor indices and $\rho$, $\rho^\prime$ are isospin indices.
$\Gamma_\mu$ is the dressed-photon+quark vertex, which can be obtained, \emph{e.g}., following Ref.\,\cite{Xu:2019ilh}.
}
\end{figure*}

Owing to the spin, flavour, and colour structure of the nucleon, the Faddeev amplitude in Eqs.\,\eqref{faveq}, \eqref{faveq3} can be decomposed as follows:
\begin{align}
	\Psi_{A B}^{C D}(p, q, P)&=
\frac{\epsilon_{rst}}{\sqrt6} \otimes
\sum_{\rho=0}^1 \psi_{\alpha\beta\gamma}^{\rho \, {\cal I}}(p, q, P) \otimes
{\mathrm F}^{\rho}_{abcd} \,,
	\label{amp}
\end{align}
where $\epsilon_{rst}$ expresses the antisymmetry of the nucleon amplitude under interchange of any two dressed-quark constituents, so the overall nucleon colour label is $u=0$.

Some additional information is useful when discussing the isospin structure.
The flavour amplitude is represented in an isospin basis spanned by the
mixed–antisymmetric (MA) and mixed–symmetric (MS) flavour combinations, rather than directly in the proton–neutron basis.
The corresponding flavour tensors are:
\begin{subequations}
\label{FlavourM}
\begin{align}
\mbox{\rm MA:} \quad F^{0}_{abcd} &= \frac{1}{\sqrt{2}}\, i \left[\sigma_2\right]_{ab} \left[\mathbb{I}\right]_{cd} \,,\\
\mbox{\rm MS:} \quad F^{1}_{abcd} &= -\frac{1}{\sqrt{6}}\, \left[\sigma_j i\sigma_2 \right]_{ab} \left[\sigma_j \right]_{cd},
\end{align}
\end{subequations}
where $\{ \sigma_i, i=1,2,3\}$ are the Pauli matrices and $[\mathbb{I} = {\rm diag}[1,1]$.
In Eq.\,\eqref{FlavourM}, under $a\leftrightarrow b$, the MA term is antisymmetric and MS is symmetric.  Interpreted in terms of potential diquark correlations \cite{Barabanov:2020jvn}, MA corresponds to the isoscalar-scalar channel and MS to isovecor-axialvector.

%
The remaining piece in Eq.\,\eqref{amp} is $\psi_{\alpha\beta\gamma}^{\rho \, {\cal I}}$, the momentum-spin component, with ${\cal I}$ recording the spin of the nucleon and $\rho = 0,1$ ranging over MA, MS.
By introducing an orthonormal set of $64$ Dirac matrix valued tensors \cite[Table~2]{Eichmann:2009en},
$\{{\mathsf X}_{\alpha \beta \gamma {\cal I}}^j(p, q, P)\}$,
\begin{align}
	\label{orth}
\frac{1}{4} {\rm tr} \,
[\overline{\mathsf X}_{\beta\alpha \delta \gamma}^i {\mathsf X}_{\alpha \beta \gamma \delta}^j]=\delta_{i j}\,,
\end{align}
this element may be decomposed into invariants as follows:
\begin{align}
\label{basise}
\psi^{\rho\, {\cal I}}_{\alpha \beta \gamma}(p, q, P)&:= \sum_{i=1}^{64}
f^{i\,\rho}(p^{2}, q^{2},z_{0},z_{1},z_{2})
{\mathsf X}_{\alpha \beta \gamma {\cal I}}^i ,
\end{align}
where the expansion coefficients, $f_{i}$, are determined by solving the Faddeev equation and depend on five scalar variables, \emph{viz}.\ with
$\hat p^2=1$,
$p_\mu^{T} = T_{\mu\nu}^P p_\mu$
etc.,
\begin{align}
&p^{2}\,, &  &q^{2}\,, &  &z_{0}=\hat{p}_{T} \cdot \hat{q}_{T}\,, &  &z_{1}=\hat{p} \cdot \hat{P}\,, &  z_{2}=\hat{q} \cdot \hat{P}\,.
\end{align}

Now inserting Eqs.\,\eqref{EqRLInteraction},  \eqref{amp} into Eq.~\eqref{faveq3c}, one obtains
\begin{equation}\label{eq:psi3-spinmom}
\begin{aligned}
\psi^{\rho\,{\mathcal I}\,(3)}_{\alpha\beta\gamma}(p,q,P)
&= \frac{2}{3}
\int_k \tilde{\mathcal G}(y)
[\gamma_\mu]_{\alpha\alpha'}S_{\alpha'\alpha''}(k_1)[\gamma_\nu]_{\beta\beta'}\\
&S_{\beta'\beta''}(\tilde k_2)
T_{\mu\nu}(k)
\psi^{\rho\,\mathcal I}_{\alpha''\beta''\gamma}(p^{(3)},q^{(3)},P).
\end{aligned}
\end{equation}
Here, the colour and flavour projections have explicitly been completed, using
\begin{equation}
[F^{\rho^\prime \, \dagger}]_{badc^\prime}\,[F^{\rho}]_{ab c d}
= \delta_{\rho^\prime \rho}\delta_{d^\prime d'}.
\end{equation}

Next, employing Eqs.\,\eqref{orth}, \eqref{basise} in processing Eq.\,\eqref{eq:psi3-spinmom}, one arrives at a system of equations for the $64\times 2$ scalar functions $\{f_i^\rho\}$ that are required to fully characterise the solution of the Faddeev equation sketched in Fig.\,\ref{FigFaddeev}.
Notably, the equations for the $f^0$ components are decoupled from those for $f^1$, but the coupling reemerges when one uses permutation symmetry to reconstruct $\psi^{\rho\,(1,2)}$.

\section{Nucleon Electromagnetic Form Factors}
\label{NCurrent}
The electromagnetic interaction current for a nucleon described by the solution of the Faddeev equation, Fig.\,\ref{FigFaddeev}, is obtained from the seven-point Schwinger function depicted in Fig.\,\ref{FigCurrent}.
For an on-shell nucleon, the general form of the current can be written as follows:
\begin{align}
J_\mu^N(Q) & = ie \Lambda_+(P_f)
[ F_1^N(Q^2) \gamma_\mu \nonumber \\
& \quad + \frac{1}{2 m_N} \sigma_{\mu\nu} Q_\nu F_2^N(Q^2) ]
\Lambda_+(P_i) \,,
\label{NucleonCurrent}
\end{align}
where $e$ is the positron charge,
the incoming and outgoing nucleon momenta are $P_{i,f}$, $P=(P_f+P_i)/2$, $Q=P_f-P_i$, $P_{i,f}^2=-m_N^2$,
$\Lambda_+(P_{i,f})$ are positive-energy nucleon-spinor projection operators,
and $F_{1,2}^N$ are the Dirac and Pauli form factors.

Following Ref.\,\cite{Sachs:1962zzc}, the nucleon charge and magnetisation distributions are normally defined as  ($\tau = Q^2/[4 m_N^2]$):
\begin{equation}
G_E^N  = F_1^N - \tau F_2^N\,,
\quad G_M^N  = F_1^N + F_2^N \,.
\label{Sachs}
\end{equation}
Magnetic moments and radii are calculated therefrom using standard definitions:
$\mu_N = G_M^N(Q^2=0)$\,;
\begin{align}
\label{StandardDef}
\langle r_F^2\rangle^N & = \left. -6 \frac{d \ln G_F^N(Q^2)}{dQ^2}\right|_{Q^2=0}\,,
\end{align}
$F=E$, $M$, except $\langle r_E^2\rangle^n = -6 G_E^{n\prime}(Q^2)|_{Q^2=0}$ because $G_E^{n}(0)=0$.

In working with the nucleon current in Fig.\,\ref{FigCurrent}, we follow the computational procedure described in Ref.\,\cite[Appendix~B]{Eichmann:2011pv}.
In that approach, the electromagnetic current is expressed in the form
\begin{equation}
J_{\mu}^N(Q):=
\sum_{a=1}^{3}
J_\mu^{aN}(Q) =
\sum_{a=1}^3\sum_{\rho^{\prime} \rho}\mathsf{F}_{\rho^{\prime} \rho}^{(a)N}[J_{\rho^{\prime} \rho}^{(a) N)}]_{\mu}\,.
\label{eq:Jcurrent}
\end{equation}
Here, owing to permutation symmetries, it is sufficient to record
\begin{equation}
\label{flavor_defF}
\mathsf{F}^{(3)N}_{\rho^\prime \rho}
=(e_N)_{d'}\,
(F^{\rho^\prime})^\dagger_{bad^\prime c^\prime }\,
Q_{c'c}\,
F^\rho_{abcd}\,
(e_N)_d,
\end{equation}
where $Q=\mathrm{diag}[q_u=2/3,q_d=-1/3]$ is the quark charge matrix associated with the photon + quark vertex,
and $e_{N=p}=(1,0)$, $e_{N=n}=(0,1)$ project onto the proton and neutron states, respectively.
Studying Eq.\,\eqref{flavor_defF}, it becomes apparent that the four elements in $\{\mathsf{F}^{(3)N}_{\rho^\prime \rho}\}$, labelled by $\{\rho^\prime, \rho\}$, are $2\times 2$ matrices in isospin space.

\begin{figure*}[t]
	\centering
	\begin{minipage}{1\textwidth}\vspace*{-0.8em}\leftline{\hspace*{1.em}\textsf{\large{A}}\hspace*{9.em}{\textsf{$\rho_{ch}^p(b)$}}\hspace*{17em}\textsf{\large{B}}\hspace*{10.75em}{\textsf{$\rho_{ch}^n(b)$}}}\end{minipage}\\
	\hspace*{-1em}\includegraphics[width=8.5cm]{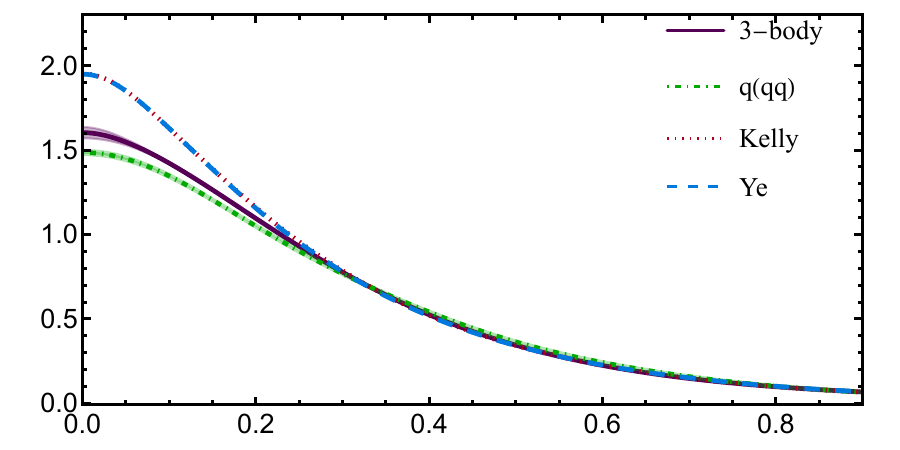}\hspace*{3em}\includegraphics[width=8.5cm]{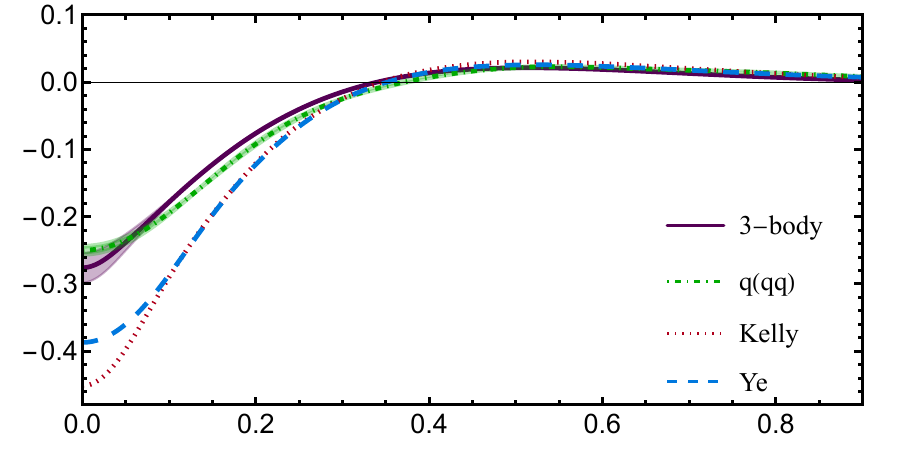}\\
	\begin{minipage}{1\textwidth}\vspace*{-0.8em}\leftline{\hspace*{11.75em}{\textsf{$b/\textrm{fm}$}}\hspace*{28em}{\textsf{$b/\textrm{fm}$}}}\end{minipage}\\
	\begin{minipage}{1\textwidth}\vspace*{0.8em}\leftline{\hspace*{1.em}\textsf{\large{C}}\hspace*{8.5em}{\textsf{$b\,\rho_{ch}^p(b)$}}\hspace*{17em}\textsf{\large{D}}\hspace*{10.75em}{\textsf{$b\,\rho_{ch}^n(b)$}}}\end{minipage}\\
	\hspace*{-1em}\includegraphics[width=8.5cm]{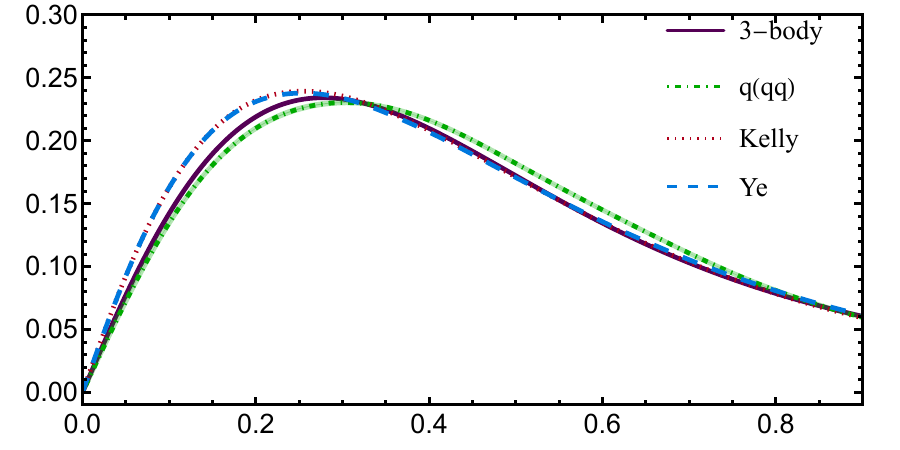}\hspace*{3em}\includegraphics[width=8.5cm]{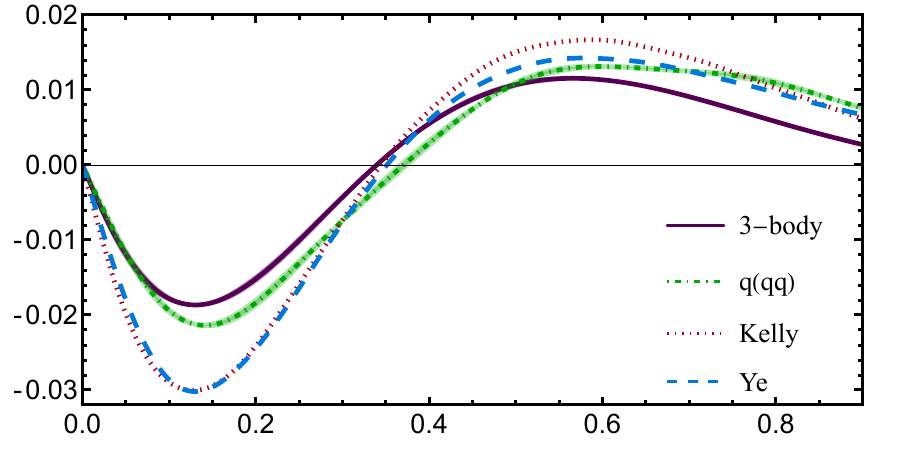}\\
	\begin{minipage}{1\textwidth}\vspace*{-0.8em}\leftline{\hspace*{11.75em}{\textsf{$b/\textrm{fm}$}}\hspace*{28em}{\textsf{$b/\textrm{fm}$}}}\end{minipage}\\
	\caption{%
Proton ($p$) and neutron ($n$) transverse charge densities.
Panel \textsf{A}(\textsf{B}) $\rho_{ch}^{p(n)}$.
Panel \textsf{C}(\textsf{D}) $b\,\rho_{ch}^{p(n)}$.
Legend.
Solid purple curve -- density obtained from form factor calculated using three-body Faddeev equation and current \cite{Yao:2024uej}, with, as therein, like-colour band indicating the associated uncertainty;
dot-dashed green -- density obtained using $q(qq)$ form factor \cite{Cheng:2025yij};
dashed blue -- density from data parametrisation in Ref.\,\cite[Ye]{Ye:2017gyb};
and dotted red -- from data parametrisation in Ref.\,\cite[Kelly]{Kelly:2004hm}.
(Distributions plotted in units of \(\text{fm}^{-2}\)  and $b=|\vec{b}|$.))\label{F1pn}	}
	\end{figure*}

\begin{figure*}[t]
	\centering
	\begin{minipage}{1\textwidth}\vspace*{-0.8em}\leftline{\hspace*{1.em}\textsf{\large{A}}\hspace*{9.em}{\textsf{$\rho_{ch}^u(b)$}}\hspace*{17em}\textsf{\large{B}}\hspace*{10.75em}{\textsf{$\rho_{ch}^d(b)$}}}\end{minipage}\\
	\hspace*{-1em}\includegraphics[width=8.5cm]{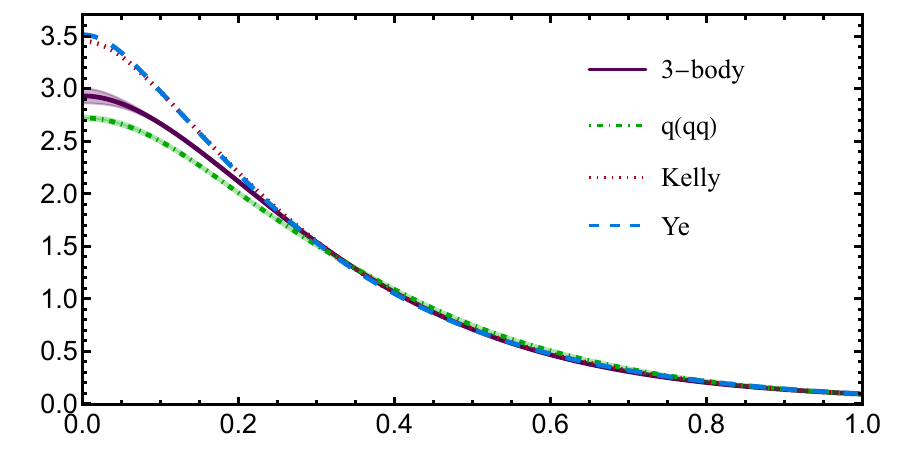}\hspace*{3em}\includegraphics[width=8.5cm]{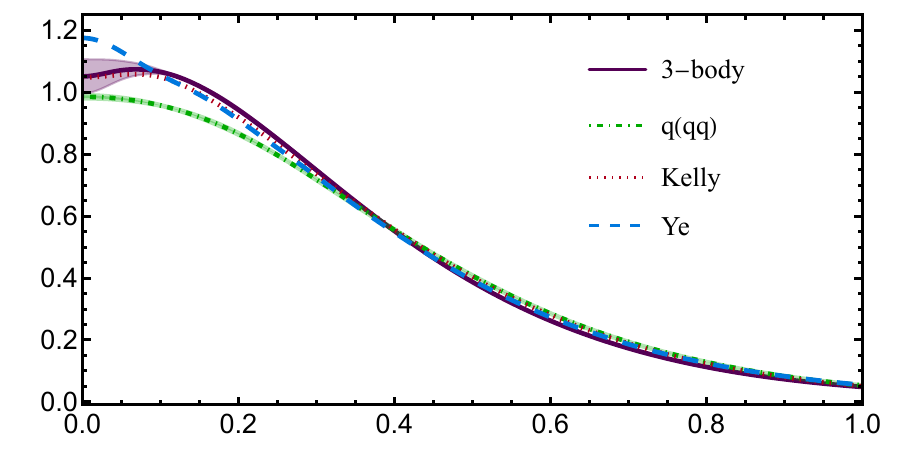}\\
	\begin{minipage}{1\textwidth}\vspace*{-0.8em}\leftline{\hspace*{12em}{\textsf{$b/\textrm{fm}$}}\hspace*{27em}{\textsf{$b/\textrm{fm}$}}}\end{minipage}\\

	\begin{minipage}{1\textwidth}\vspace*{0.8em}\leftline{\hspace*{1.em}\textsf{\large{C}}\hspace*{9.em}{\textsf{$b\,\rho_{ch}^u(b)$}}\hspace*{17em}\textsf{\large{D}}\hspace*{10.75em}{\textsf{$b\,\rho_{ch}^d(b)$}}}\end{minipage}\\
	\hspace*{-1em}\includegraphics[width=8.5cm]{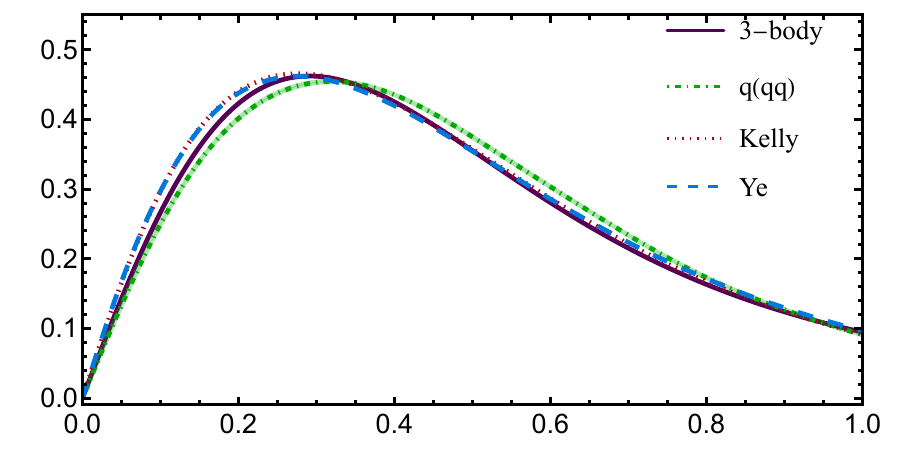}\hspace*{3em}\includegraphics[width=8.5cm]{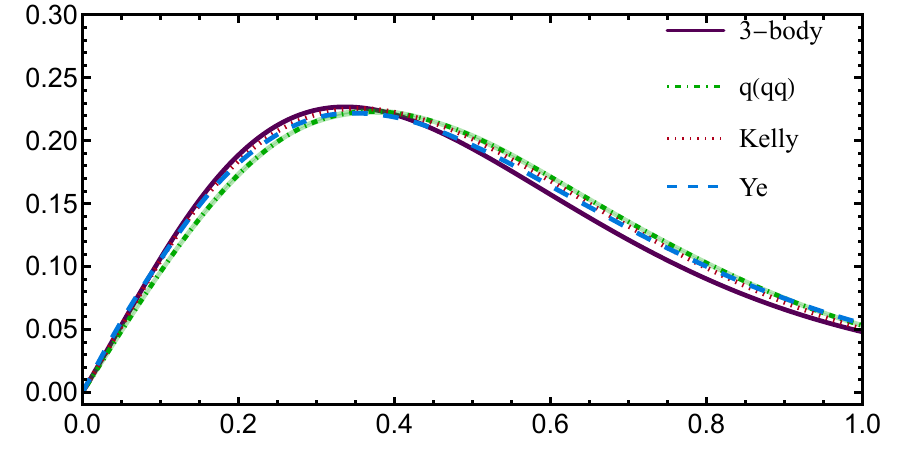}\\
	\begin{minipage}{1\textwidth}\vspace*{-0.8em}\leftline{\hspace*{12em}{\textsf{$b/\textrm{fm}$}}\hspace*{27em}{\textsf{$b/\textrm{fm}$}}}\end{minipage}\\
	\caption{%
$u$- and $d$-quark transverse charge densities.
Panel \textsf{A}(\textsf{B}) $\rho_{ch}^{u(d)}$.
Panel \textsf{C}(\textsf{D}) $b\,\rho_{ch}^{u(d)}$.
Legend.
Solid purple curve -- density obtained from form factor calculated using three-body Faddeev equation and current \cite{Yao:2024uej}, with, as therein, like-colour band indicating the associated uncertainty;
dot-dashed green -- density obtained using $q(qq)$ form factor \cite{Cheng:2025yij};
dashed blue -- density from data parametrisation in Ref.\,\cite[Ye]{Ye:2017gyb};
and dotted red -- from data parametrisation in Ref.\,\cite[Kelly]{Kelly:2004hm}.
(Distributions plotted in units of \(\text{fm}^{-2}\) and $b=|\vec{b}|$.)
	\label{F1ud}}
\end{figure*}

\begin{figure*}[t]
	\centering
	\begin{minipage}{1\textwidth}\vspace*{-0.8em}\leftline{\hspace*{1.em}\textsf{\large{A}}\hspace*{9.em}{\textsf{${\rho}_{m}^p/\kappa_p$}}\hspace*{16em}\textsf{\large{B}}\hspace*{10.75em}{\textsf{${\rho}_{m}^n/\kappa_n$}}}\end{minipage}\\
	\hspace*{-1em}\includegraphics[width=8.5cm]{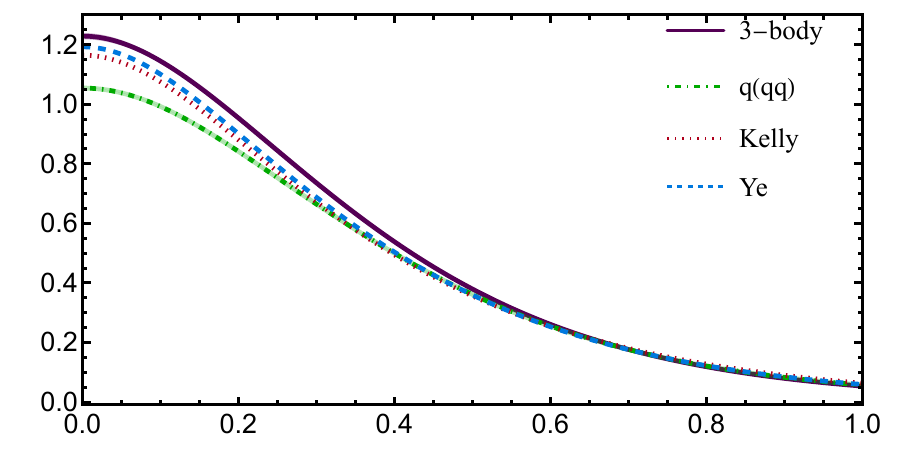}\hspace*{3em}\includegraphics[width=8.5cm]{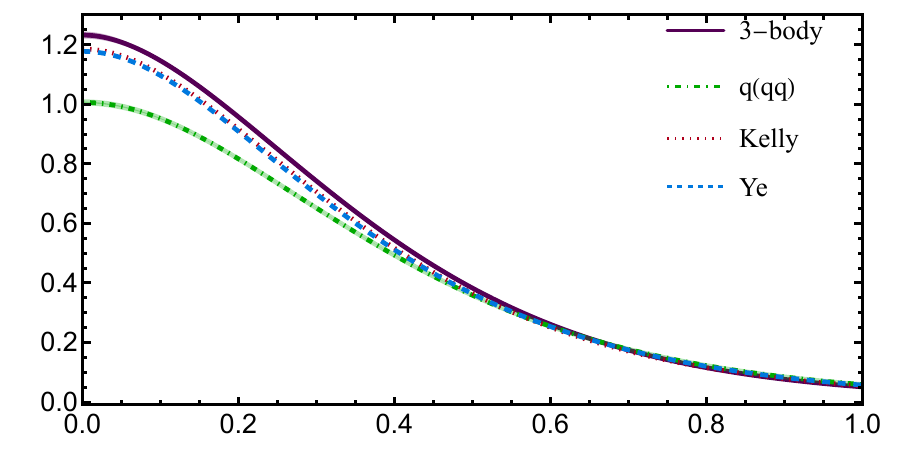}\\
	\begin{minipage}{1\textwidth}\vspace*{-0.8em}\leftline{\hspace*{12em}{\textsf{$b/\textrm{fm}$}}\hspace*{27em}{\textsf{$b/\textrm{fm}$}}}\end{minipage}\\
	\begin{minipage}{1\textwidth}\vspace*{0.8em}\leftline{\hspace*{1.em}\textsf{\large{C}}\hspace*{9.em}{\textsf{$b\,{\rho}_{m}^p/\kappa_p$}}\hspace*{15.5em}\textsf{\large{D}}\hspace*{10.75em}{\textsf{$b\,{\rho}_{M}^n/\kappa_n$}}}\end{minipage}\\
	\hspace*{-1em}\includegraphics[width=8.5cm]{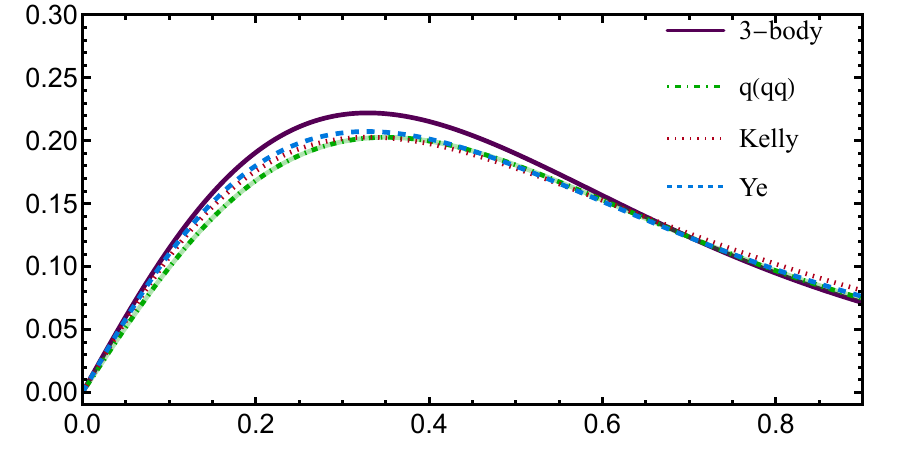}\hspace*{3em}\includegraphics[width=8.5cm]{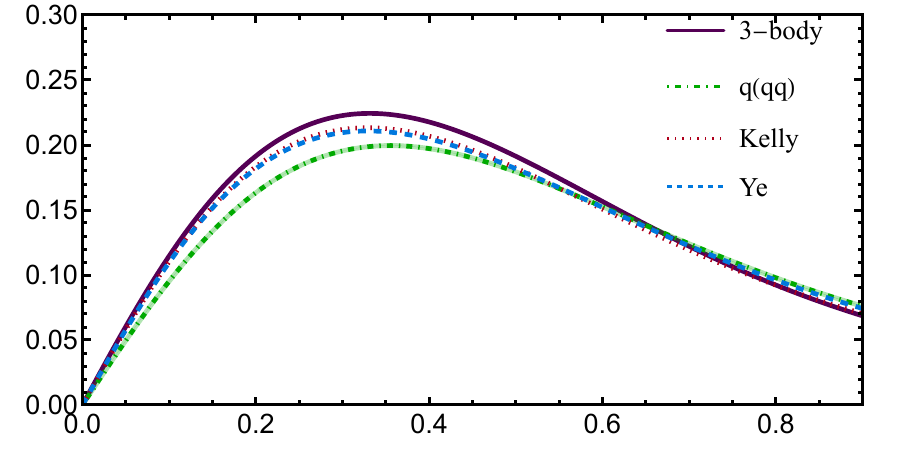}\\
	\begin{minipage}{1\textwidth}\vspace*{-0.8em}\leftline{\hspace*{12em}{\textsf{$b/\textrm{fm}$}}\hspace*{27em}{\textsf{$b/\textrm{fm}$}}}\end{minipage}\\
	\caption{%
Proton ($p$) and neutron ($n$) transverse magnetisation densities, Eq.\,\eqref{Tmagnet}.
Panel \textsf{A}(\textsf{B}) $\rho_{m}^{p(n)}$.
Panel \textsf{C}(\textsf{D}) $b\,\rho_{m}^{p(n)}$.
Legend.
Solid purple curve -- density obtained from form factor calculated using three-body Faddeev equation and current \cite{Yao:2024uej}, with, as therein, like-colour band indicating the associated uncertainty;
dot-dashed green -- density obtained using $q(qq)$ form factor \cite{Cheng:2025yij};
dashed blue -- density from data parametrisation in Ref.\,\cite[Ye]{Ye:2017gyb};
and dotted red -- from data parametrisation in Ref.\,\cite[Kelly]{Kelly:2004hm}.
(Distributions plotted in units of \(\text{fm}^{-2}\) and $b=|\vec{b}|$).)
	\label{fig:F2pn}}
	\end{figure*}

\begin{figure*}[t]
	\centering
	\begin{minipage}{1\textwidth}\vspace*{-0.8em}\leftline{\hspace*{1.em}\textsf{\large{A}}\hspace*{9.em}{\textsf{${\rho}_{m}^u/\kappa_u$}}\hspace*{16em}\textsf{\large{B}}\hspace*{10.75em}{\textsf{${\rho}_{m}^d/\kappa_d$}}}\end{minipage}\\
	\hspace*{-1em}\includegraphics[width=8.5cm]{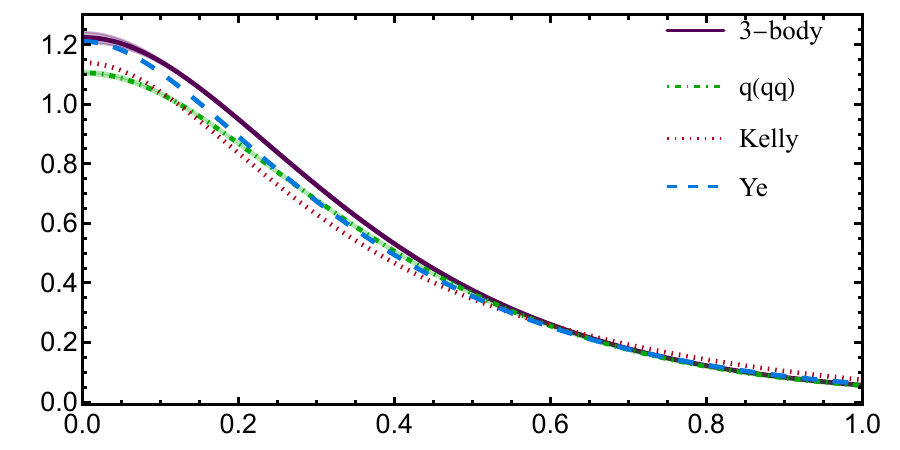}\hspace*{3em}\includegraphics[width=8.5cm]{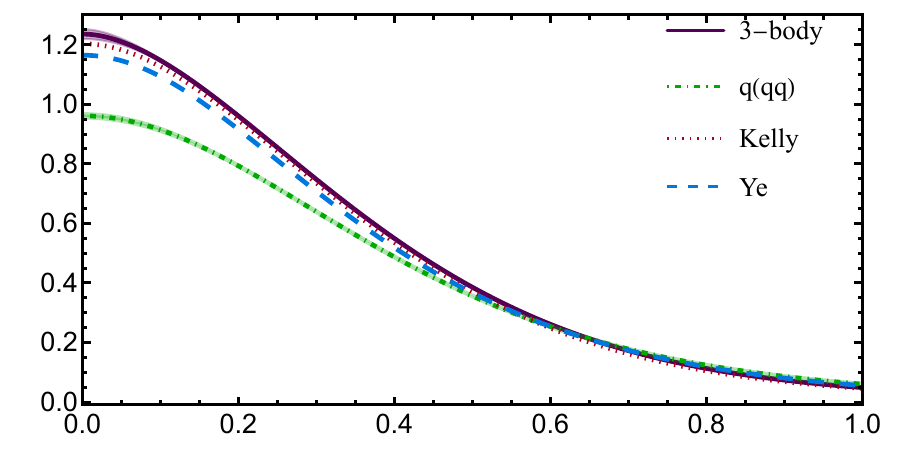}\\
	\begin{minipage}{1\textwidth}\vspace*{-0.8em}\leftline{\hspace*{12em}{\textsf{$b/\textrm{fm}$}}\hspace*{27em}{\textsf{$b/\textrm{fm}$}}}\end{minipage}\\
	\begin{minipage}{1\textwidth}\vspace*{0.8em}\leftline{\hspace*{1.em}\textsf{\large{C}}\hspace*{9.em}{\textsf{$b\,{\rho}_{m}^u/\kappa_u$}}\hspace*{15.5em}\textsf{\large{D}}\hspace*{10.75em}{\textsf{$b\,{\rho}_{m}^d/\kappa_d$}}}\end{minipage}\\
	\hspace*{-1em}\includegraphics[width=8.5cm]{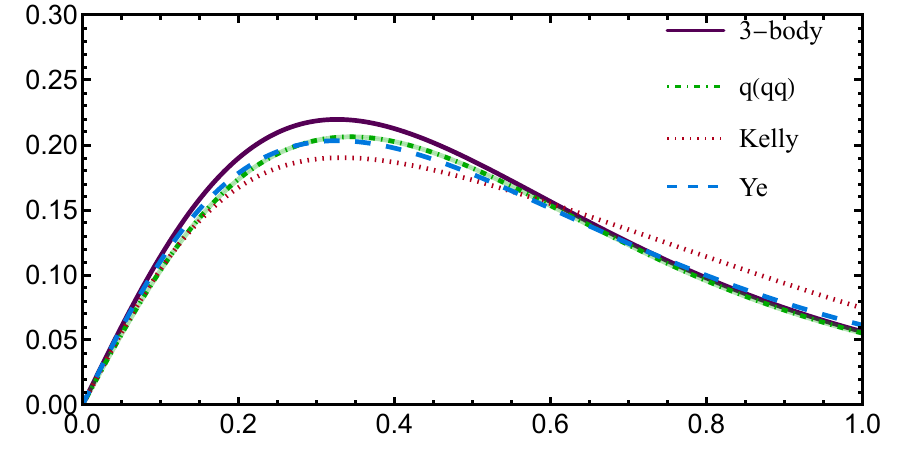}\hspace*{3em}\includegraphics[width=8.5cm]{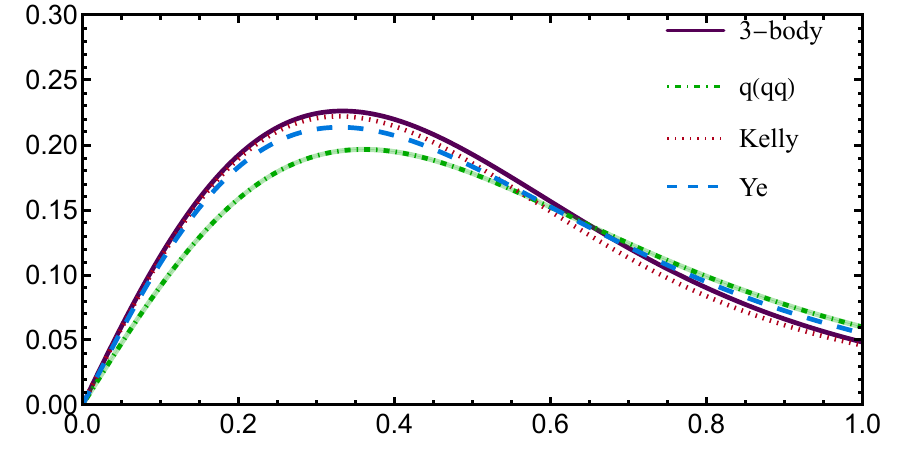}\\
	\begin{minipage}{1\textwidth}\vspace*{-0.8em}\leftline{\hspace*{12em}{\textsf{$b/\textrm{fm}$}}\hspace*{27em}{\textsf{$b/\textrm{fm}$}}}\end{minipage}\\
	\caption{%
$u$- and $d$-quark transverse magnetisation densities, Eq.\,\eqref{Tmagnet}.
Panel \textsf{A}(\textsf{B}) $\rho_{m}^{u(d)}$.
Panel \textsf{C}(\textsf{D}) $b\,\rho_{m}^{u(d)}$.
Legend.
Solid purple curve -- density obtained from form factor calculated using three-body Faddeev equation and current \cite{Yao:2024uej}, with, as therein, like-colour band indicating the associated uncertainty;
dot-dashed green -- density obtained using $q(qq)$ form factor \cite{Cheng:2025yij};
dashed blue -- density from data parametrisation in Ref.\,\cite[Ye]{Ye:2017gyb};
and dotted red -- from data parametrisation in Ref.\,\cite[Kelly]{Kelly:2004hm}.
(Distributions plotted in units of \(\text{fm}^{-2}\) and $b=|\vec{b}|$).)
	\label{fig:F2ud}}
\end{figure*}

The physical proton and neutron currents (hence, their form factors) are obtained by combining the $\rho=0,1$ contributions via the flavour matrices \(\mathsf{F}_{\rho^\prime \rho}^{(a)}\) in Eq.~\eqref{eq:Jcurrent}.  Evaluating the expressions in Eq.\,\eqref{flavor_defF}, one finds:
\begin{equation}\label{flavor_p}
\mathsf{F}^{(3)p}  =
\begin{pmatrix}
  \tfrac{2}{3} & 0\\
  0 & 0
\end{pmatrix},
\qquad
\mathsf{F}^{(3)n} =
\begin{pmatrix}
 -\tfrac{1}{3} & 0\\
  0 & \tfrac{1}{3}
\end{pmatrix}.
\end{equation}

It remains only to report the mathematical expression for the photon coupling to the quark in Fig.\,\ref{FigCurrent}:
\begin{align}
\label{ImFF}
 [&J_{\rho^{\prime} \rho}^{(3)}]_{\mu}^{\delta^{\prime} \delta}
  =\int_{dp dq}\bar{\psi}^{\rho^\prime \delta^{\prime} }_{\beta^{\prime} \alpha^{\prime} \gamma^{\prime}} \left(p_f, q_f, P_f\right)
 S_{\alpha^{\prime} \alpha}\left(p_1\right) S_{\beta^{\prime} \beta}\left(p_2\right)\nonumber \\
&\times [\chi_{\mu}(p_3,Q)]_{\gamma^{\prime} \gamma} \left[\psi^{\rho \delta }_{\alpha \beta \gamma}(p_i, q_i, P_i)
- \psi^{\rho \delta (3)}_{\alpha \beta \gamma}(p_i, q_i, P_i)\right],
\end{align}
where $p_i, q_i$ and $p_f, q_f$ are the incoming and outgoing relative momenta of the bound-state valence dressed quarks;
$P_i$, $P_f$ are the incoming, outgoing nucleon momenta;
and $p_1,\ p_2$, $p_3^{ \pm}=p_3 \pm Q / 2$ are the quark momenta.
For $\xi=1/3$ in Eq.\,\eqref{Involvesxi}, one has
\begin{subequations}
\label{NewMomenta}
\begin{align}
p_i & = p-Q/3\,, \quad p_f= p+Q/3\,,\quad q_f=q_i=q\,, \\
p_1 & = -q-\frac{p}{2}+\frac{P}{3}\,, \quad
p_2 = q-\frac{p}{2}+\frac{P}{3}\,, \\
P_{i}& =P-Q/2\,, \quad P_f=P+Q/2\,. \label{DefPif}
\end{align}
\end{subequations}
We note that Eq.\,\eqref{DefPif} merely specifies a Breit frame calculation; hence, is unconnected with the value of $\xi$.

Equation~\eqref{ImFF} also involves the unamputated photon + quark vertex, $\chi_{\mu}(p_3,Q)=S(p_3^{+}) \Gamma_{\mu}(p,Q) S(p_3^{-})$, wherein the amputated vertex, $\Gamma_{\mu}(p,Q)$, is obtained as the solution of an inhomogeneous Bethe-Salpeter equation.  This equation and its solution are discussed extensively elsewhere \cite{Xu:2019ilh, Xu:2021mju}.

Using Eqs.\,\eqref{eq:Jcurrent}, \eqref{flavor_p} and permutation symmetry, one arrives at the following nucleon electromagnetic currents:
\begin{subequations}
\label{CurrentNucleon}
\begin{align}
\mathrm{proton:}\quad  & J^{\delta^{\prime} \delta ; p }_{\mu} =[2 J_{00}^{(3)}]^{\delta^{\prime} \delta}_{\mu},\\
\mathrm{neutron:} \quad  &J^{\delta^{\prime} \delta; n}_{\mu} =[J_{11}^{(3)}-J_{00}^{(3)}]^{\delta^{\prime} \delta}_{\mu}.
\end{align}
\end{subequations}
Referring to Eqs.\,\eqref{FlavourM}, \eqref{flavor_defF}, then Eq.\,\eqref{CurrentNucleon} means that there is no contribution to charged spin-half baryon form factors from MS pieces of the bound-state amplitude.
A similar result is described elsewhere in the context of quark + diquark pictures of nucleon structure \cite[Appendix\,C.1]{Wilson:2011aa}:
for such a baryon, ``elastic'' scattering from the axialvector diquark piece of its Faddeev wave functions does not contribute to its electromagnetic form factors.

The calculation described herein is made with reference to a hadron scale, $\zeta_{\cal H}$, whereat all properties of a given hadron are carried by its dressed valence degrees of freedom \cite{Yin:2023dbw}.
Hence, flavour-separated currents may be defined as follows:
\begin{subequations}
\label{FlavourSep}
\begin{align}
\mathrm{u\, in\,proton:}\quad  &
2 J^{p }_{\mu} + J^{n  }_{\mu}
=3 [J_{00}^{(3)}]_\mu + [J_{11}^{(3)}]_\mu \,, \\
\mathrm{d\, in\,proton:}\quad  & J^{p }_{\mu} + 2 J^{n  }_{\mu}
= 2 [J_{11}^{(3)}]_\mu \,.
\end{align}
\end{subequations}
These results entail that the in-proton $d$-quark form factors receive no contribution from MA components of the proton Faddeev amplitude.
Again, this has a long-known analogue in the quark + diquark picture of nucleon structure; namely, $d$ quarks cannot be perceived by a hard probe unless axialvector diquark correlations are included in the proton Faddeev wave function \cite{Close:1988br, Cui:2021gzg, Cheng:2023kmt}.

With explicit expressions for the proton and neutron currents in hand, one can compute the associated Sachs form factors (the trace is over Dirac indices):
\begin{subequations}
\label{eq:GEGM}
\begin{align}
G_E^N(Q^2) &=
\frac{1}{2i\sqrt{1+\tau}}\,
{\rm tr} \left[J^{N}_\mu \hat{P}_{\mu}\right], \\
G_M^N(Q^2) &= \frac{i}{4\tau}\,
{\rm tr}\left[J^{N}_\mu \gamma_{\mu}^{T}\right].
\end{align}
\end{subequations}
From these expressions, one readily obtains the nucleon Dirac and Pauli form factors via Eq.\,\eqref{Sachs}.
The flavour-separated forms of these functions then follow from Eq.\,\eqref{FlavourSep}.

Canonical normalisation of the nucleon Faddeev amplitude is fixed  by ensuring $G_E^p(0)=1$.
Current conservation, guaranteed by our formulation of the interaction current, is expressed in, amongst other things, the result $G_E^{n}(0)\equiv0$.

\section{\label{S:density}Charge and Magnetisation Densities}
\subsection{Transverse Dirac charge densities}
Consider first nucleon transverse charge densities:
\begin{subequations}
\label{Tcharge}
\begin{align}
\rho_{ch}^{\mathpzc t}(|\vec{b}|) & = \int\frac{d^2\Delta}{(2\pi)^2} \,{\rm e}^{i \vec{b}\cdot \vec{\Delta}}
F_1^{\mathpzc t}(\Delta^2) \\
& = \int_0^\infty \frac{d |\vec{\Delta}|}{2\pi} \, |\vec{\Delta}|\, J_0(|\vec{b}||\vec{\Delta}|)
F_1^{\mathpzc t}(\Delta^2) \,.
\end{align}
\end{subequations}
Here, $\mathpzc t=p,n,u,d$, \emph{i.e}., proton, neutron, $u$-quark, $d$-quark, with -- see Eq.\,\eqref{FlavourSep}:
\begin{subequations}
\label{FlavourDensities}
\begin{align}
\rho_{ch}^u(|\vec{b}|) & = 2 \rho_{ch}^p(|\vec{b}|) + \rho_{ch}^n(|\vec{b}|)\,,\\
\rho_{ch}^d(|\vec{b}|) & = \rho_{ch}^p(|\vec{b}|) + 2 \rho_{ch}^n(|\vec{b}|)\,.
\end{align}
\end{subequations}
Further,
$J_0$ is a Bessel function of the first kind;
$\vec{b} = (b_x,b_y)=|\vec{b}| (\cos\phi_b,\sin\phi_b)$;
and $Q=(\vec{\Delta},0,0)$.

Figure~\ref{F1pn} depicts transverse charge densities for the proton and neutron.
The primary theory curves, labelled ``3-body'',  were obtained from the form factors reported in Ref.\,\cite{Yao:2024uej}, which were calculated using the formalism described in Sects.\,\ref{S:Schwinger}, \ref{NCurrent}.
These curves have an associated uncertainty.
It expresses a $1\sigma$-band deriving from the approach used to produce form factors at large-$\Delta^2$, namely, the Schlessinger point method (SPM) \cite{Schlessinger:1966zz, PhysRev.167.1411, Tripolt:2016cya}.
Details are provided in Ref.\,\cite{Yao:2024uej}: evidently, the impact on our results of this uncertainty is practically negligible.

The second set of theory predictions was obtained using the quark + interacting-diquark, $q(qq)$, picture of nucleon structure described in Ref.\,\cite{Cheng:2025yij}.
As first revealed, independently, in Refs.\,\cite{Cahill:1988dx, Reinhardt:1989rw, Efimov:1990uz}, the $q(qq)$ picture emerges when one exploits the pairing tendency of fermions to simplify the Faddeev equation sketched in Fig.\,\ref{FigFaddeev}.
One thereby arrives at a bound-state equation that describes dressed quarks and fully-interacting quark-quark (diquark) correlations built therefrom, which bind together into a baryon, at least in part, because of the continual exchange of roles between the spectator and diquark-participant quarks.
Many effectual studies of nucleon properties have employed this $q(qq)$ scheme; see, \emph{e.g}., Refs.\,\cite{Barabanov:2020jvn, Roberts:2020hiw, Ding:2022ows, Cheng:2023kmt, Chen:2023zhh, Yu:2024ovn, Eichmann:2025tzm, Yu:2025fer, Cheng:2025yij}.

A significant merit of the dynamical $q(qq)$ approach is that it greatly simplifies the form of the Poincar\'e-covariant amplitude required to describe the nucleon: instead of the 128 independent tensor structures needed in the fully $3$-body treatment, only $8$ are required in the $q(qq)$ approximation. Hence, it is much easier to use and may deliver additional insights. 
The limitation of the Ref.\,\cite{Cheng:2025yij} $q(qq)$ approach is its less rigorous connection with QCD dynamics: rather than being directly calculated, QCD-constrained \emph{Ans\"atze} are introduced for dressed-constituent propagators and interaction vertices.
One aim of the comparisons presented herein is to assist in judging the fidelity of the \emph{Ans\"atze} used.

The form factors in Ref.\,\cite{Cheng:2025yij} are reported as functions of $x=Q^2/m_N^2$.
To obtain results in GeV$^2$, we employed $m_N=0.94\,$GeV.

In all cases, there is fair agreement between the 3-body predictions and those obtained using $q(qq)$ input.
Moreover, both sets of predictions agree semiquantitatively with the results obtained using fits to data as input to Eq.\,\eqref{Tcharge}.
Those (modest) differences between theory and data-based inferences which are apparent might be connected with the omission of meson-baryon final state interactions (meson cloud effects) in the theoretical calculations employed herein.
Further studies are needed before that can be confirmed.

The proton transverse charge density is positive definite on the entire $|\vec{b}|$ domain.
Thus, one must expect that at every point in the light-front transverse plane, the valence $u$-quark density exceeds the valence $d$-quark density.

On the other hand, the neutron transverse charge density is negative on a sizeable neighbourhood of the neutron's centre of transverse momentum (CoTM, $|\vec{b}|=0$).
It then becomes positive at $|\vec{b}|\approx 0.35\,$fm and remains so thereafter.
These features were noted previously \cite{Miller:2007uy}.
Mathematically, they are consequences of:
(\emph{a}) $F_1^n(\Delta^2=0)=0$;
(\emph{b}) the non-positive character of $F_1^n(\Delta^2)$ and $\Delta^2$-dependent profile of this Dirac form factor, namely, in magnitude, with increasing $\Delta^2$, it rises to a maximum and then falls monotonically to zero;
and (c) the oscillatory behaviour of the weight $|\vec{\Delta}|\, J_0(|\vec{b}||\vec{\Delta}|)$ at fixed $|\vec{b}|$.

In explanation, we note that, on $|\vec{b}|\simeq 0$, the $\rho_{ch}^n(|\vec{b}|)$ integrand is zero at $|\vec{\Delta}|=0$,
negative until its first zero,
then oscillates with large steps between adjacent zeros and ever decreasing peak-height magnitudes; consequently, the value of the integral is dominated by the negative support at low $|\vec{\Delta}|=0$, with all subsequent negative peaks giving larger negative contributions to the integral than the smaller positive peaks.
As $|\vec{b}|$ increases, there comes a value at which the peak-height magnitude of the integrand's first positive peak exceeds that of the first negative peak, which always occurs at a lower value of $|\vec{\Delta}|$.
In this case, the positive contribution dominates on the domain bounded by the integrand's second nontrivial zero; and this pattern is repeated for all subsequent oscillations.
Thus, the integral is positive.
This pattern remains in effect thereafter, so $\rho_{ch}^n(|\vec{b}|)$ remains positive on $|\vec{b}|\gtrsim 0.35\,$fm.

Considering the above discussion, then one must expect that on $|\vec{b}|\lesssim 0.35\,$fm in the light-front transverse plane, the $u$-quark density in the proton actually exceeds twice the $d$-quark density.
This feature does not persist on the complementary domain; see, also, Ref.\,\cite{Miller:2007uy}.

Having made this observation, transverse densities for the $u$- and $d$-quarks in the proton are displayed in Fig.\,\ref{F1ud}.
Evidently, the expectations described above are realised.
Moreover, both in-proton valence-quark light-front transverse charge densities are positive definite.


\begin{table}[b]
\caption{Theory predictions for in-proton valence-quark transverse density radii, defined via Eq.\,\eqref{TransverseRadii} with an obvious analogue for the magnetisation, in which the anomalous dressed valence-quark magnetic moment must be divided out.  Inferences from data are also listed.
\label{tab:rad}}
\begin{tabular*}
{\hsize}
{
l@{\extracolsep{0ptplus1fil}}
|l@{\extracolsep{0ptplus1fil}}
l@{\extracolsep{0ptplus1fil}}
l@{\extracolsep{0ptplus1fil}}
l@{\extracolsep{0ptplus1fil}}}\hline\hline
\centering
& \(\lambda_1^u\)/fm & \(\lambda_1^d\)/fm & \(\lambda_2^u\)/fm & \(\lambda_2^d\)/fm \\
\hline
3-body$\ $  & 0.68 & 0.69 & 0.63 &0.68 \\
q(qq)   & 0.53   &0.57 & 0.62   &0.67 \\
\cite[Ye]{Ye:2017gyb}     & 0.66(1)  & 0.67(1) & 0.71(12)  & 0.69(14) \\
\cite[Kelly]{Kelly:2004hm} $\ $ & 0.65   & 0.66&  0.68  &0.76\\
\hline\hline
\end{tabular*}
\end{table}

To quantify some of the remarks made above, one may consider root-mean-square radii connected with the transverse-plane Dirac charge densities, \emph{viz}.\ $\lambda_1^{u,d}$, where
\begin{align}
\label{TransverseRadii}
(\lambda_1^{u,d})^2 & := \frac{\int\! d^2 \vec b\, |\vec b|^2 \rho_1^{u,d}(|\vec b|)}{\int\! d^2\vec b\, \rho_1^f(|\vec b|)}   \nonumber \\
& =\left. - 4 \frac{d}{dQ^2} \ln F_1^{u,d}(Q^2)\right|_{Q^2=0}\,.
\end{align}
Predictions for these radii are compared with inferences from parametrisations of data in Table~\ref{tab:rad}.
The $3$-body transverse-plane Dirac radii are noticeably larger than those obtained using $q(qq)$ input.
Whilst not readily apparent in Fig.\,\ref{F1ud}, this outcome owes to the $3$-body densities having greater support on $|\vec{b}|\gtrsim 1.1\,$fm.
(Equally, referring to Ref.\,\cite{Cheng:2025yij}, the $Q^2\simeq 0$ slope of the $3$-body form factors is greater than that obtained in the $q(qq)$ approach.)
Turning to a comparison with inferences from data -- see Table~\ref{tab:rad}, the $3$-body predictions are in better agreement.
Overall, we judge that the valence $u$- and $d$-quark transverse-plane Dirac radii are practically indistinguishable.

\subsection{Transverse Pauli magnetisation densities}
By analogy with Eq.\,\eqref{Tcharge}, one may define light-front transverse magnetisation densities:
\begin{equation}
\rho_{m}^{\mathpzc t}(|\vec{b}|)
 = \int_0^\infty \frac{d |\vec{\Delta}|}{2\pi} \, |\vec{\Delta}|\, J_0(|\vec{b}||\vec{\Delta}|)
F_2^{\mathpzc t}(\Delta^2) \,.
\label{Tmagnet}
\end{equation}
With this definition, one recovers the anomalous magnetic moments via:
\begin{equation}
\label{kappdef}
\kappa^{\mathpzc t} = 2\pi \int_0^\infty d |\vec{b}| \, |\vec{b}|\,\rho_m^{\mathpzc t}(|\vec{b}|)\,.
\end{equation}

Theory predictions for $\kappa^{\mathpzc t} $ are compared with empirical determinations in Table~\ref{tab:kappa}.
The results obtained using the $3$-body approach described above deliver underestimated values.
As explained in Ref.\,\cite{Yao:2024uej}, that is a failing of RL truncation, which yields a photon + quark vertex whose dressed-quark anomalous magnetic moment term is too small.
This weakness is corrected in higher-order truncations \cite{Chang:2010hb}.
Such corrections have been implemented in studies of mesons \cite{Xu:2022kng}.  In future, it may be possible to adapt this scheme to baryons.

\begin{table}[b]
\caption{Theory predictions for anomalous magnetic moments, \(\kappa\) in Eq.\,\eqref{kappdef}, compared with empirical determinations \cite{ParticleDataGroup:2024cfk} and using Eq.\,\eqref{FlavourDensities}, where necessary.
\label{tab:kappa}}
\begin{tabular*}
{\hsize}
{
l@{\extracolsep{0ptplus1fil}}
|c@{\extracolsep{0ptplus1fil}}
c@{\extracolsep{0ptplus1fil}}
c@{\extracolsep{0ptplus1fil}}
cl@{\extracolsep{0ptplus1fil}}}\hline\hline
\centering
& \(\kappa_p\) & \(\kappa_n\) & \(\kappa_u\) & \(\kappa_d\) \\
\hline
Exp.     & 1.75   & $-1.91$ & 1.59   & $-2.08$ \\
3-body $\ $  & 1.23   & $-1.33$ & 1.13   & $-1.43$ \\
q(qq)   & 1.80   & $-1.86$ & 1.74   & $-1.92$ \\
\hline\hline
\end{tabular*}
\end{table}

Transverse magnetisation densities for the proton and neutron are displayed in Fig.\,\ref{fig:F2pn}.
In this case, all densities are positive definite;
and the $3$-body predictions agree best with empirical inferences (the anomalous magnetic moment has been divided out of $\rho_m^{p,n}$), although the $q(qq)$ results are nonetheless a fair match.

\begin{figure*}[t]
\centering
\begin{minipage}{1\textwidth}\vspace*{1em}\leftline{\hspace*{-1.5em}\textsf{\large{3-body\ A}}\hspace*{7.em}{\textsf{$\rho_{T,ch,\tilde m}^p(b_x=0,b_y)$}}\hspace*{13.em}\textsf{\large{B}}\hspace*{7em}{\textsf{$\rho_{T,ch,\tilde m}^n(b_x=0,b_y)$}}}\end{minipage}\\
\hspace*{-1em}\includegraphics[width=8.5cm]{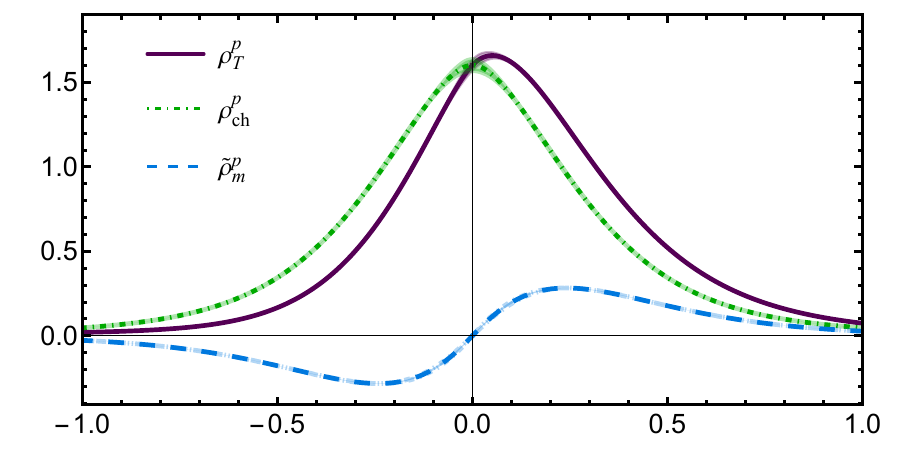}\hspace*{3em}\includegraphics[width=8.5cm]{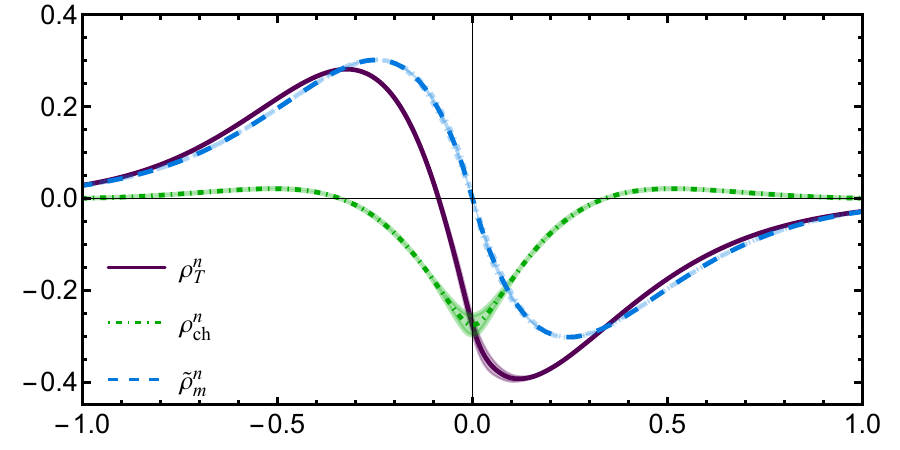}\\
\begin{minipage}{1\textwidth}\vspace*{-0.8em}\leftline{\hspace*{12em}{\textsf{$b_y/\textrm{fm}$}}\hspace*{27em}{\textsf{$b_y/\textrm{fm}$}}}\end{minipage}

\begin{minipage}{1\textwidth}\vspace*{1em}\leftline{\hspace*{-1.5em}\textsf{\large{q(qq)\ \ C}}\hspace*{7.em}{\textsf{$\rho_{T,ch,\tilde m}^p(b_x=0,b_y)$}}\hspace*{13.em}\textsf{\large{D}}\hspace*{7em}{\textsf{$\rho_{T,ch,\tilde m}^n(b_x=0,b_y)$}}}\end{minipage}\\
\hspace*{-1em}\includegraphics[width=8.5cm]{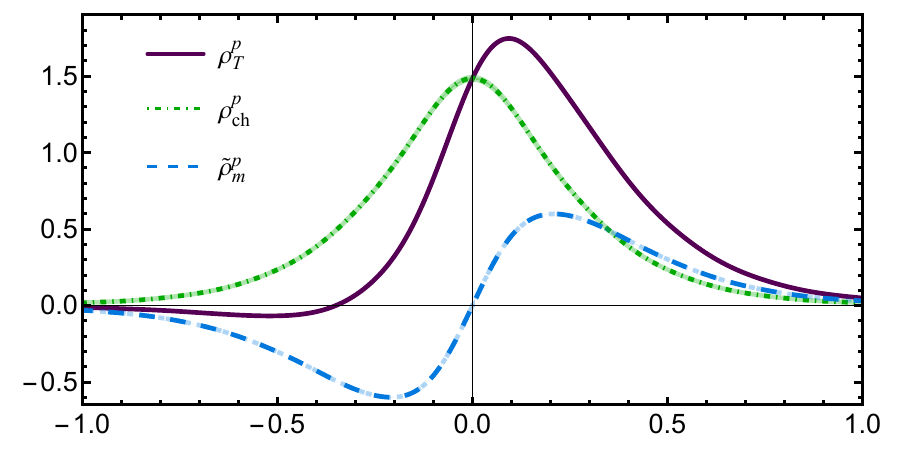}\hspace*{3em}\includegraphics[width=8.5cm]{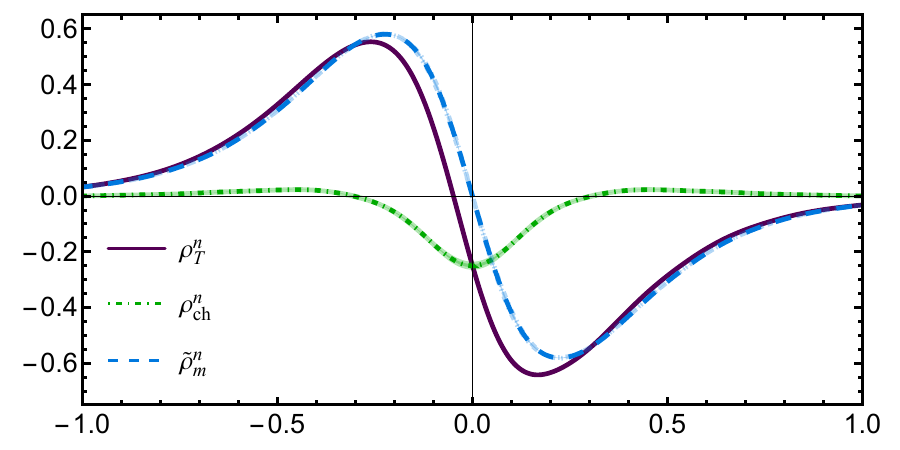}\\
\begin{minipage}{1\textwidth}\vspace*{-0.8em}\leftline{\hspace*{12em}{\textsf{$b_y/\textrm{fm}$}}\hspace*{27em}{\textsf{$b_y/\textrm{fm}$}}}\end{minipage}
\caption{All three proton and neutron light-front-transverse densities for a nucleon with transverse polarisation along $+\hat x$.
Panel \textsf{A}(\textsf{B}): $\rho^{p(n)}$ from three-body Faddeev equation \cite{Yao:2024uej}.
Panel \textsf{C}(\textsf{D}): $\rho^{p(n)}$ from q(qq) treatment of nucleon \cite{Cheng:2025yij}.
Legend.
Solid purple curve -- $\rho_T^{p(n)}$;
dot-dashed green -- $\rho_{ch}^{p(n)}$;
dashed blue -- $\tilde\rho_{m}^{p(n)}$.
Like coloured bands display SPM-related $1\sigma$ uncertainties.  They are practically negligible.
(Density distributions plotted in units of \(\text{fm}^{-2}\)).}
\label{fig:F1pnT}
\end{figure*}

\begin{figure*}[t]
\centering
\begin{minipage}{1\textwidth}\vspace*{1em}\leftline{\hspace*{-1.5em}\textsf{\large{3-body\ A}}\hspace*{7.em}{\textsf{$\rho_{T,ch,\tilde m}^u(b_x=0,b_y)$}}\hspace*{13.em}\textsf{\large{B}}\hspace*{7em}{\textsf{$\rho_{T,ch,\tilde m}^d(b_x=0,b_y)$}}}\end{minipage}\\
\hspace*{-1em}\includegraphics[width=8.5cm]{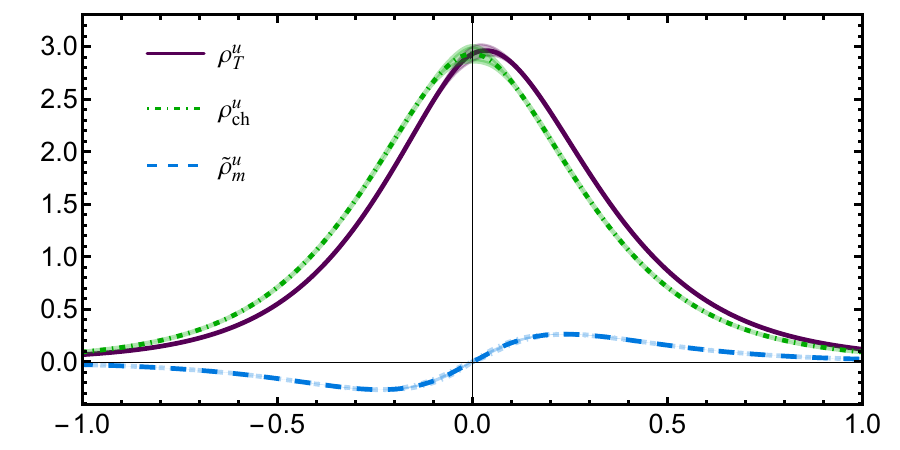}\hspace*{3em}\includegraphics[width=8.5cm]{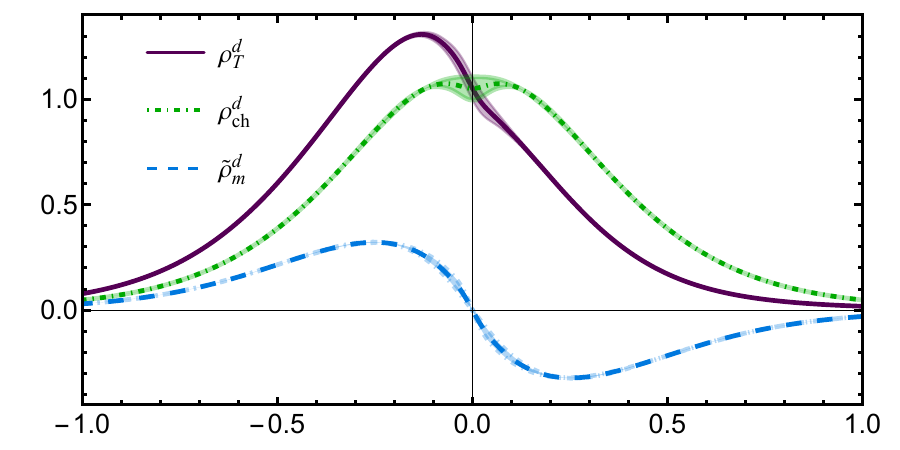}\\
\begin{minipage}{1\textwidth}\vspace*{-0.8em}\leftline{\hspace*{12em}{\textsf{$b_y/\textrm{fm}$}}\hspace*{27em}{\textsf{$b_y/\textrm{fm}$}}}\end{minipage}

\begin{minipage}{1\textwidth}\vspace*{1em}\leftline{\hspace*{-1.5em}\textsf{\large{q(qq)\ \ C}}\hspace*{7.em}{\textsf{$\rho_{T,ch,\tilde m}^u(b_x=0,b_y)$}}\hspace*{13.em}\textsf{\large{D}}\hspace*{7em}{\textsf{$\rho_{T,ch,\tilde m}^d(b_x=0,b_y)$}}}\end{minipage}\\
\hspace*{-1em}\includegraphics[width=8.5cm]{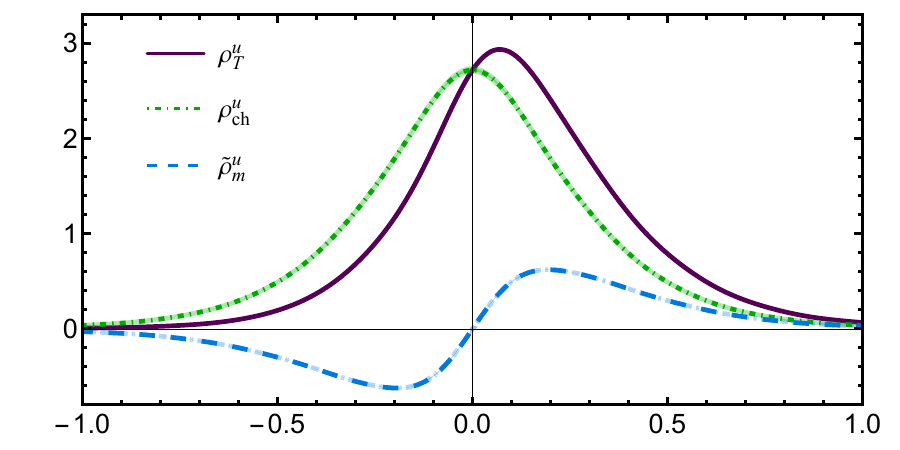}\hspace*{3em}\includegraphics[width=8.5cm]{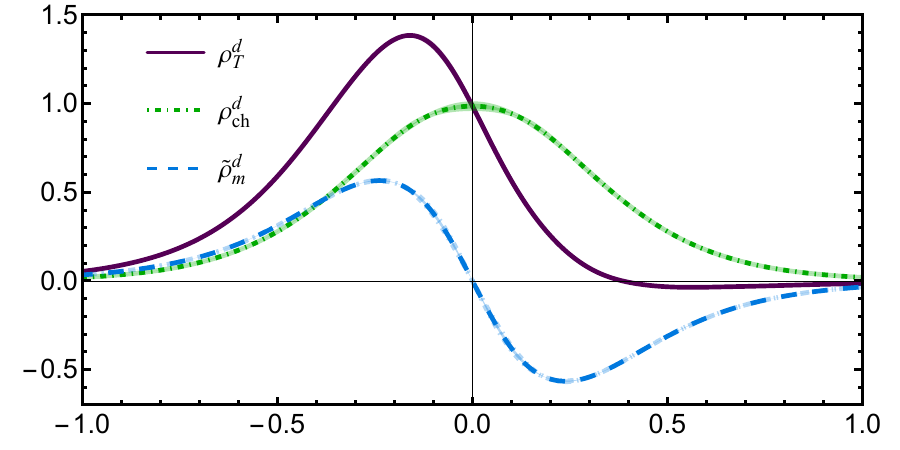}\\
\begin{minipage}{1\textwidth}\vspace*{-0.8em}\leftline{\hspace*{12em}{\textsf{$b_y/\textrm{fm}$}}\hspace*{27em}{\textsf{$b_y/\textrm{fm}$}}}\end{minipage}
\caption{Flavour separation of light-front-transverse densities for a proton with transverse polarisation along $+\hat x$.
Panel \textsf{A}(\textsf{B}): $\rho^{u(d)}$ from three-body Faddeev equation \cite{Yao:2024uej}.
Panel \textsf{C}(\textsf{D}): $\rho^{u(d)}$ from q(qq) treatment of nucleon \cite{Cheng:2025yij}.
Legend.
Solid purple curve -- $\rho_T^{u(d)}$;
dot-dashed green -- $\rho_{ch}^{u(d)}$;
dashed blue -- $\tilde\rho_{m}^{u(d)}$.
Once again, like coloured bands display SPM-related $1\sigma$ uncertainties.  They are practically negligible.
(Density distributions plotted in units of \(\text{fm}^{-2}\)).}
\label{fig:F1udT}
\end{figure*}

\begin{figure*}[t]
\centering
\begin{minipage}{1\textwidth}\vspace*{1em}\leftline{\hspace*{-1.5em}\textsf{\large{A\ proton}}\hspace*{7.em}{\textsf{$\phantom{\rho_{T,ch,\tilde m}^u(b_x=0,b_y)}$}}\hspace*{13.em}\textsf{\large{B\ neutron}}\hspace*{7em}{\textsf{$\phantom{\rho_{T,ch,\tilde m}^d(b_x=0,b_y)}$}}}\end{minipage}\\
\hspace*{-1em}\includegraphics[width=8.5cm]{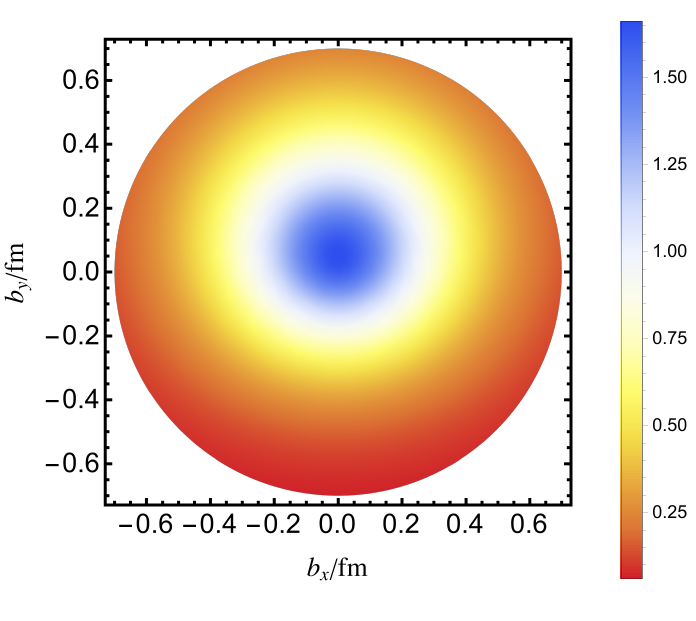}\hspace*{3em}\includegraphics[width=8.5cm]{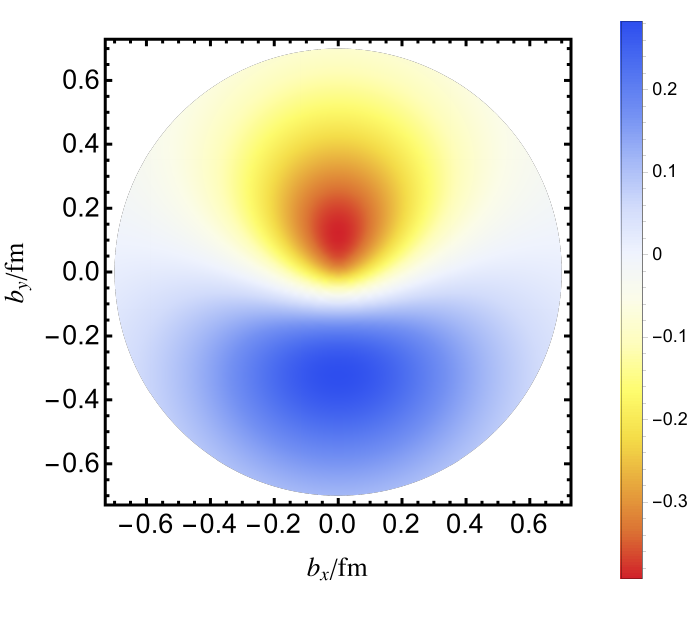}\\
%
\caption{
With polarisation in the $+ b_x$ direction, two-dimensional transverse charge densities of the proton (Panel~\textsf{A}) and neutron (Panel~\textsf{B}).
Both results obtained using the $3$-body framework.
Images for the $q(qq)$ picture are practically equivalent.
(The unit of the density distribution is \(\text{fm}^{-2}\).)
\label{fig:F1pnT3d}}
\end{figure*}

The associated in-proton flavour-separated magnetisation densities are depicted in Fig.\,\ref{fig:F2ud}:
owing to division by $\kappa$, each is positive definite.
As observed elsewhere \cite{Cheng:2025yij}, the profiles in these panels highlight that the valence $d$-quark is magnetically very active within the proton: in magnitude, the $d$-quark profile is approximately equal (although opposite in sign) to the $u$-quark profile, even though the proton contains two valence $u$-quarks and only one $d$-quark.
This outcome is correlated with suggestions from the $q(qq)$ picture that, at $\zeta_{\cal H}$, the $u$ valence quark angular momentum in the proton is much less than that of the $d$ valence quark \cite{Cheng:2023kmt, Yu:2024qsd, Yu:2024ovn}.
Furthermore, since, in a scalar diquark only proton, the $d$ quark is locked within a magnetically inert $[ud]$ correlation, the presence of axialvector diquarks in the proton is crucial to understanding $\rho_m^u$.
Translated into the $3$-body framework, this becomes a statement that MS components of the Faddeev wave function are important to obtain a valid description of in-proton valence $d$-quark behaviour.

Predictions for transverse-plane Pauli radii are recorded in Table~\ref{tab:rad}.
Defined by analogy with Eq.\,\eqref{TransverseRadii}, with the second line amended via normalisation by the $\kappa^{u,d}$ values, they are compared in the table with results inferred from fits to data.
In this case, the $3$-body and $q(qq)$ predictions agree, with ordering that is consistent with experimental inferences.
Hence, one may conclude, that the in-proton transverse-plane valence $d$-quark Pauli radius is $9(2)$\% larger than the analogous valence $u$-quark radius.

\subsection{Transverse charge densities for transversely polarised nucleon}
For a nucleon prepared with polarisation transverse to its direction of motion, the light-front transverse charge density receives an additional contribution \cite{Burkardt:2002hr, Carlson:2007xd}:
\begin{align}
\label{SpinChargeDensity1}
\rho_{T}(|\vec{b}|) = \rho_{\text{ch}}(|\vec{b}|) + \sin(\phi_b - \phi_s) \frac{1}{2m_N |\vec{b}|} \tilde\rho_m(|\vec{b}|),
\end{align}
where $\vec{S} = (\cos \phi_s , \sin \phi_s )$ is the nucleon polarisation unit-vector and
\begin{subequations}
\label{SpinChargeDensity}
\begin{align}
\tilde\rho_m^{\mathpzc t}(|\vec{b}|) &
=- |\vec{b}| \frac{\partial {\rho}_M^{\mathpzc t}(|\vec{b}|)}
{\partial |\vec{b}|} \\
& = |\vec{b}| \int_0^\infty\frac{d |\vec{\Delta}| }{2\pi} \Delta^2 J_1(|\vec{\Delta}||\vec{b}|)
F_2^{\mathpzc t}(\Delta^2).
\end{align}
\end{subequations}
(\emph{N.B}.\ The relative sign in Eq.\,\eqref{SpinChargeDensity1} reflects the remarks in Ref.\,\cite[Addendum]{Burkardt:2002hr}.)

As illustrated in Fig.\,\ref{IMF}, associating the nucleon spin direction with the $+\hat x$-axis ($\phi_s = 0$), then the light-front transverse charge density is distorted in the $\hat y$ direction.
On any domain for which $\tilde\rho_m^{\mathpzc t}(|\vec{b}|) >0$, the shift is an enhancement in
$\rho_{\text{ch}}(|\vec{b}|)$ on the positive $\hat y$ axis ($\phi_b=\pi/2$) and a depletion on the negative $\hat y$ axis.

The light-front transverse charge densities for a $+\hat x$ polarised nucleon are depicted in Fig.\,\ref{fig:F1pnT}.
The $3$-body and $q(qq)$ results are qualitatively and semi-quantitatively identical.
For the polarised nucleons, there is a depletion of the charge density in the $-\hat y$ direction, with that strength shifted to the $+\hat y$ domain.
This means that, in the presence of polarisation, the positive and negative charge densities are not symmetric around $b_y=0$.
Instead, positive charge is shifted to positive values of $b_y$ and negative charge is shifted to negative $b_y$.

Flavour-separated images for a transversely polarised proton are displayed in Fig.\,\ref{fig:F1udT}.
As anticipated above, on $+\hat y$, the $u$-quark densities are enhanced, whereas they are depleted on $-\hat y$.
The reverse is true for the $d$-quark densities and the magnitude of the displacements is larger.  The latter owes to the larger magnitude of $\kappa_d$ -- see Table~\ref{tab:kappa} -- which leads to a magnification of the shift generated by the second term in Eq.\,\eqref{SpinChargeDensity1}.

In Fig.\,\ref{fig:F1pnT3d}, we draw two-dimensional density maps of the transverse charge densities for transversely polarised nucleons obtained using the $3$-body approach.
Based on the above discussion, those obtained in the $q(qq)$ picture are qualitatively equivalent; so, we don't plot them.
Supposing one is working in the infinite momentum frame, then these images depict a nucleon moving toward the observer.
Plainly, the distribution is not rotationally symmetric around the proton direction of motion; instead, the positive charge density is distorted in the $+\hat y$ direction.
Reflection symmetry through the $\hat y$ axis is preserved.
Although the effects for proton and neutron are similar, owing to its positive/negative character, the neutron's positive-charge density drift into the $+\hat y$ direction throughout the transverse plane is, perhaps, most noticeable.

\section{\label{S:S}Summary and Perspective}
Using input from $3$-body and $q(qq)$ Faddeev equation results for nucleon elastic electromagnetic form factors, we computed an array of light-front transverse densities for the proton and neutron and their dressed valence-quark constituents, \emph{i.e}., flavour separations of the proton and neutron results.
In completing such calculations, it was crucial that both frameworks are Poincar\'e-covariant and deliver form factor predictions that extend to very large values of spacelike $Q^2$.
All results are collected and discussed in Sec.\,\ref{S:density}.

The key point is that the two complementary pictures of nucleon structure deliver predictions in qualitative and near-quantitative agreement, and that these predictions match expectations based on modern parametrisations of available data, where such are available.
Underlying this success is the expression of emergent hadron mass (EHM) in the quark + quark interaction that drives both the $3$-body and $q(qq)$ results.

Looking to the future, one may anticipate a time at which lattice-QCD results reach a level of reliability that will allow quantitative comparisons with the full array of our predictions to be made.
Given the common Schwinger function foundation of both continuum and lattice approaches, agreement would lend additional support to the EHM paradigm that underlies the continuum predictions.

On another track, the results presented herein could be extended to nucleon-to-resonance transition transverse densities.
Such analyses are underway within the $q(qq)$ framework.
They will become possible in the $3$-body approach once the Poincar\'e invariant transition form factors have been calculated, an effort which is in train.

\begin{acknowledgments}
We thank C.~Lorc\'e for pointing us to the erratum of Ref.\,\cite{Burkardt:2002hr}.
%
Work supported by:
National Natural Science Foundation of China, grant no.\ 12135007;
Natural Science Foundation of Anhui Pro\-vince (grant no.\ 2408085QA028);
Helmholtz-Zentrum Dresden-Rossendorf, under the High Potential
Programme;
%
Ministerio Espa\~nol de Ciencia, Innovaci\'on y Universidades (MICINN) grant no.\ PID2022-140440NB-C22;
and Junta de Andaluc\'{\i}a contract nos.\ PAIDI FQM-370 and PCI+D+i under the title: ``Tecnolog\'{\i}as avanzadas para la exploraci\'on del universo y sus componentes" (Code AST22-0001).
\end{acknowledgments}



\begin{thebibliography}{87}
\providecommand{\natexlab}[1]{#1}
\providecommand{\url}[1]{\texttt{#1}}
\providecommand{\urlprefix}{URL }
\expandafter\ifx\csname urlstyle\endcsname\relax
  \providecommand{\doi}[1]{doi:\discretionary{}{}{}#1}\else
  \providecommand{\doi}[1]{doi:\discretionary{}{}{}\begingroup
  \urlstyle{rm}\url{#1}\endgroup}\fi
\providecommand{\bibinfo}[2]{#2}

\bibitem[{Hofstadter(1956)}]{Hofstadter:1956qs}
\bibinfo{author}{R.~Hofstadter}, \bibinfo{title}{{Electron scattering and
  nuclear structure}}, \bibinfo{journal}{Rev. Mod. Phys.} \bibinfo{volume}{28}
  (\bibinfo{year}{1956}) \bibinfo{pages}{214--254}.

\bibitem[{Perdrisat et~al.(2007)Perdrisat, Punjabi, and
  Vanderhaeghen}]{Perdrisat:2006hj}
\bibinfo{author}{C.~F. Perdrisat}, \bibinfo{author}{V.~Punjabi},
  \bibinfo{author}{M.~Vanderhaeghen}, \bibinfo{title}{{Nucleon electromagnetic
  form factors}}, \bibinfo{journal}{Prog. Part. Nucl. Phys.}
  \bibinfo{volume}{59} (\bibinfo{year}{2007}) \bibinfo{pages}{694--764}.

\bibitem[{Punjabi et~al.(2015)Punjabi, Perdrisat, Jones, Brash, and
  Carlson}]{Punjabi:2015bba}
\bibinfo{author}{V.~Punjabi}, \bibinfo{author}{C.~F. Perdrisat},
  \bibinfo{author}{M.~K. Jones}, \bibinfo{author}{E.~J. Brash},
  \bibinfo{author}{C.~E. Carlson}, \bibinfo{title}{{The Structure of the
  Nucleon: Elastic Electromagnetic Form Factors}}, \bibinfo{journal}{Eur. Phys.
  J. A} \bibinfo{volume}{51} (\bibinfo{year}{2015}) \bibinfo{pages}{79}.

\bibitem[{Gao and Vanderhaeghen(2022)}]{Gao:2021sml}
\bibinfo{author}{H.~Gao}, \bibinfo{author}{M.~Vanderhaeghen},
  \bibinfo{title}{{The proton charge radius}}, \bibinfo{journal}{Rev. Mod.
  Phys.} \bibinfo{volume}{94}~(\bibinfo{number}{1}) (\bibinfo{year}{2022})
  \bibinfo{pages}{015002}.

\bibitem[{Cui et~al.(2022{\natexlab{a}})Cui, Binosi, Roberts, and
  Schmidt}]{Cui:2022fyr}
\bibinfo{author}{Z.-F. Cui}, \bibinfo{author}{D.~Binosi},
  \bibinfo{author}{C.~D. Roberts}, \bibinfo{author}{S.~M. Schmidt},
  \bibinfo{title}{{Hadron and light nucleus radii from electron scattering}},
  \bibinfo{journal}{Chin. Phys. C} \bibinfo{volume}{46}~(\bibinfo{number}{12})
  (\bibinfo{year}{2022}{\natexlab{a}}) \bibinfo{pages}{122001}.

\bibitem[{Gilfoyle(2018)}]{Gilfoyle:2018xsa}
\bibinfo{author}{G.~Gilfoyle}, \bibinfo{title}{{Future Measurements of the
  Nucleon Elastic Electromagnetic Form Factors at Jefferson Lab}},
  \bibinfo{journal}{EPJ Web Conf.} \bibinfo{volume}{172} (\bibinfo{year}{2018})
  \bibinfo{pages}{02004}.

\bibitem[{Wojtsekhowski(2020)}]{Wojtsekhowski:2020tlo}
\bibinfo{author}{B.~Wojtsekhowski}, \bibinfo{title}{{Flavor Decomposition of
  Nucleon Form Factors -- arXiv:2001.02190 [nucl-ex]}}, \bibinfo{year}{2020}.

\bibitem[{Yamazaki et~al.(2009)Yamazaki, Aoki, Blum, Lin, Ohta, Sasaki,
  Tweedie, and Zanotti}]{Yamazaki:2009zq}
\bibinfo{author}{T.~Yamazaki}, \bibinfo{author}{Y.~Aoki},
  \bibinfo{author}{T.~Blum}, \bibinfo{author}{H.-W. Lin},
  \bibinfo{author}{S.~Ohta}, \bibinfo{author}{S.~Sasaki},
  \bibinfo{author}{R.~Tweedie}, \bibinfo{author}{J.~Zanotti},
  \bibinfo{title}{{Nucleon form factors with 2+1 flavor dynamical domain-wall
  fermions}}, \bibinfo{journal}{Phys. Rev. D} \bibinfo{volume}{79}
  (\bibinfo{year}{2009}) \bibinfo{pages}{114505}.

\bibitem[{Bhattacharya et~al.(2014)Bhattacharya, Cohen, Gupta, Joseph, Lin, and
  Yoon}]{Bhattacharya:2013ehc}
\bibinfo{author}{T.~Bhattacharya}, \bibinfo{author}{S.~D. Cohen},
  \bibinfo{author}{R.~Gupta}, \bibinfo{author}{A.~Joseph},
  \bibinfo{author}{H.-W. Lin}, \bibinfo{author}{B.~Yoon},
  \bibinfo{title}{{Nucleon Charges and Electromagnetic Form Factors from
  2+1+1-Flavor Lattice QCD}}, \bibinfo{journal}{Phys. Rev. D}
  \bibinfo{volume}{89}~(\bibinfo{number}{9}) (\bibinfo{year}{2014})
  \bibinfo{pages}{094502}.

\bibitem[{Chambers et~al.(2017)}]{QCDSF:2017ssq}
\bibinfo{author}{A.~J. Chambers}, et~al., \bibinfo{title}{{Electromagnetic form
  factors at large momenta from lattice QCD}}, \bibinfo{journal}{Phys. Rev. D}
  \bibinfo{volume}{96}~(\bibinfo{number}{11}) (\bibinfo{year}{2017})
  \bibinfo{pages}{114509}.

\bibitem[{Alexandrou et~al.(2019)Alexandrou, Bacchio, Constantinou, Finkenrath,
  Hadjiyiannakou, Jansen, Koutsou, and Vaquero
  Aviles-Casco}]{Alexandrou:2018sjm}
\bibinfo{author}{C.~Alexandrou}, \bibinfo{author}{S.~Bacchio},
  \bibinfo{author}{M.~Constantinou}, \bibinfo{author}{J.~Finkenrath},
  \bibinfo{author}{K.~Hadjiyiannakou}, \bibinfo{author}{K.~Jansen},
  \bibinfo{author}{G.~Koutsou}, \bibinfo{author}{A.~Vaquero Aviles-Casco},
  \bibinfo{title}{{Proton and neutron electromagnetic form factors from lattice
  QCD}}, \bibinfo{journal}{Phys. Rev. D}
  \bibinfo{volume}{100}~(\bibinfo{number}{1}) (\bibinfo{year}{2019})
  \bibinfo{pages}{014509}.

\bibitem[{Alexandrou et~al.(2023)Alexandrou, Bacchio, Constantinou, Finkenrath,
  Hadjiyiannakou, Jansen, Koutsou, and Spanoudes}]{Alexandrou:2023qor}
\bibinfo{author}{C.~Alexandrou}, \bibinfo{author}{S.~Bacchio},
  \bibinfo{author}{M.~Constantinou}, \bibinfo{author}{J.~Finkenrath},
  \bibinfo{author}{K.~Hadjiyiannakou}, \bibinfo{author}{K.~Jansen},
  \bibinfo{author}{G.~Koutsou}, \bibinfo{author}{G.~Spanoudes},
  \bibinfo{title}{{Nucleon electromagnetic form factors using $N_f$=2+1+1
  twisted mass fermion ensembles at the physical mass point}},
  \bibinfo{journal}{PoS} \bibinfo{volume}{LATTICE2022} (\bibinfo{year}{2023})
  \bibinfo{pages}{114}.

\bibitem[{Syritsyn et~al.(2025)Syritsyn, Engelhardt, Krieg, Negele, and
  Pochinsky}]{Syritsyn:2025fiu}
\bibinfo{author}{S.~Syritsyn}, \bibinfo{author}{M.~Engelhardt},
  \bibinfo{author}{S.~Krieg}, \bibinfo{author}{J.~Negele},
  \bibinfo{author}{A.~Pochinsky}, \bibinfo{title}{{Nucleon electromagnetic form
  factors at large momentum from Lattice QCD}}, \bibinfo{journal}{PoS}
  \bibinfo{volume}{LATTICE2024} (\bibinfo{year}{2025}) \bibinfo{pages}{340}.

\bibitem[{de~Melo et~al.(2009)de~Melo, Frederico, Pace, Pisano, and
  Salme}]{deMelo:2008rj}
\bibinfo{author}{J.~P. B.~C. de~Melo}, \bibinfo{author}{T.~Frederico},
  \bibinfo{author}{E.~Pace}, \bibinfo{author}{S.~Pisano},
  \bibinfo{author}{G.~Salme}, \bibinfo{title}{{Time- and Spacelike Nucleon
  Electromagnetic Form Factors beyond Relativistic Constituent Quark Models}},
  \bibinfo{journal}{Phys. Lett. B} \bibinfo{volume}{671} (\bibinfo{year}{2009})
  \bibinfo{pages}{153--157}.

\bibitem[{Giannini and Santopinto(2015)}]{Giannini:2015zia}
\bibinfo{author}{M.~M. Giannini}, \bibinfo{author}{E.~Santopinto},
  \bibinfo{title}{{The hypercentral Constituent Quark Model and its application
  to baryon properties}}, \bibinfo{journal}{Chin. J. Phys.}
  \bibinfo{volume}{53} (\bibinfo{year}{2015}) \bibinfo{pages}{020301}.

\bibitem[{Sufian et~al.(2017)Sufian, de~T\'eramond, Brodsky, Deur, and
  Dosch}]{Sufian:2016hwn}
\bibinfo{author}{R.~S. Sufian}, \bibinfo{author}{G.~F. de~T\'eramond},
  \bibinfo{author}{S.~J. Brodsky}, \bibinfo{author}{A.~Deur},
  \bibinfo{author}{H.~G. Dosch}, \bibinfo{title}{{Analysis of nucleon
  electromagnetic form factors from light-front holographic QCD : The spacelike
  region}}, \bibinfo{journal}{Phys. Rev. D}
  \bibinfo{volume}{95}~(\bibinfo{number}{1}) (\bibinfo{year}{2017})
  \bibinfo{pages}{014011}.

\bibitem[{Mondal et~al.(2020)Mondal, Xu, Lan, Zhao, Li, Chakrabarti, and
  Vary}]{Mondal:2019jdg}
\bibinfo{author}{C.~Mondal}, \bibinfo{author}{S.~Xu}, \bibinfo{author}{J.~Lan},
  \bibinfo{author}{X.~Zhao}, \bibinfo{author}{Y.~Li},
  \bibinfo{author}{D.~Chakrabarti}, \bibinfo{author}{J.~P. Vary},
  \bibinfo{title}{{Proton structure from a light-front Hamiltonian}},
  \bibinfo{journal}{Phys. Rev. D} \bibinfo{volume}{102}~(\bibinfo{number}{1})
  (\bibinfo{year}{2020}) \bibinfo{pages}{016008}.

\bibitem[{Ahmady et~al.(2021)Ahmady, Chakrabarti, Mondal, and
  Sandapen}]{Ahmady:2021qed}
\bibinfo{author}{M.~Ahmady}, \bibinfo{author}{D.~Chakrabarti},
  \bibinfo{author}{C.~Mondal}, \bibinfo{author}{R.~Sandapen},
  \bibinfo{title}{{Nucleon electroweak form factors using spin-improved
  holographic light-front wavefunctions}}, \bibinfo{journal}{Nucl. Phys. A}
  \bibinfo{volume}{1016} (\bibinfo{year}{2021}) \bibinfo{pages}{122334}.

\bibitem[{Xu et~al.(2021{\natexlab{a}})Xu, Mondal, Lan, Zhao, Li, and
  Vary}]{Xu:2021wwj}
\bibinfo{author}{S.~Xu}, \bibinfo{author}{C.~Mondal}, \bibinfo{author}{J.~Lan},
  \bibinfo{author}{X.~Zhao}, \bibinfo{author}{Y.~Li}, \bibinfo{author}{J.~P.
  Vary}, \bibinfo{title}{{Nucleon structure from basis light-front
  quantization}}, \bibinfo{journal}{Phys. Rev. D}
  \bibinfo{volume}{104}~(\bibinfo{number}{9})
  (\bibinfo{year}{2021}{\natexlab{a}}) \bibinfo{pages}{094036}.

\bibitem[{Eichmann et~al.(2016)Eichmann, Sanchis-Alepuz, Williams, Alkofer, and
  Fischer}]{Eichmann:2016yit}
\bibinfo{author}{G.~Eichmann}, \bibinfo{author}{H.~Sanchis-Alepuz},
  \bibinfo{author}{R.~Williams}, \bibinfo{author}{R.~Alkofer},
  \bibinfo{author}{C.~S. Fischer}, \bibinfo{title}{{Baryons as relativistic
  three-quark bound states}}, \bibinfo{journal}{Prog. Part. Nucl. Phys.}
  \bibinfo{volume}{91} (\bibinfo{year}{2016}) \bibinfo{pages}{1--100}.

\bibitem[{Burkert and Roberts(2019)}]{Burkert:2019bhp}
\bibinfo{author}{V.~D. Burkert}, \bibinfo{author}{C.~D. Roberts},
  \bibinfo{title}{{Colloquium: Roper resonance: Toward a solution to the
  fifty-year puzzle}}, \bibinfo{journal}{Rev. Mod. Phys.} \bibinfo{volume}{91}
  (\bibinfo{year}{2019}) \bibinfo{pages}{011003}.

\bibitem[{Binosi(2022)}]{Binosi:2022djx}
\bibinfo{author}{D.~Binosi}, \bibinfo{title}{{Emergent Hadron Mass in Strong
  Dynamics}}, \bibinfo{journal}{Few Body Syst.}
  \bibinfo{volume}{63}~(\bibinfo{number}{2}) (\bibinfo{year}{2022})
  \bibinfo{pages}{42}.

\bibitem[{Ding et~al.(2023)Ding, Roberts, and Schmidt}]{Ding:2022ows}
\bibinfo{author}{M.~Ding}, \bibinfo{author}{C.~D. Roberts},
  \bibinfo{author}{S.~M. Schmidt}, \bibinfo{title}{{Emergence of Hadron Mass
  and Structure}}, \bibinfo{journal}{Particles}
  \bibinfo{volume}{6}~(\bibinfo{number}{1}) (\bibinfo{year}{2023})
  \bibinfo{pages}{57--120}.

\bibitem[{Ferreira and Papavassiliou(2023)}]{Ferreira:2023fva}
\bibinfo{author}{M.~N. Ferreira}, \bibinfo{author}{J.~Papavassiliou},
  \bibinfo{title}{{Gauge Sector Dynamics in QCD}}, \bibinfo{journal}{Particles}
  \bibinfo{volume}{6}~(\bibinfo{number}{1}) (\bibinfo{year}{2023})
  \bibinfo{pages}{312--363}.

\bibitem[{Achenbach et~al.(2025)Achenbach, Carman, Gothe, Joo, Mokeev, and
  Roberts}]{Achenbach:2025kfx}
\bibinfo{author}{P.~Achenbach}, \bibinfo{author}{D.~S. Carman},
  \bibinfo{author}{R.~W. Gothe}, \bibinfo{author}{K.~Joo},
  \bibinfo{author}{V.~I. Mokeev}, \bibinfo{author}{C.~D. Roberts},
  \bibinfo{title}{{Electroexcitation of Nucleon Resonances and the Emergence of
  Hadron Mass}}, \bibinfo{journal}{Symmetry}
  \bibinfo{volume}{17}~(\bibinfo{number}{7}) (\bibinfo{year}{2025})
  \bibinfo{pages}{1106}.

\bibitem[{Eichmann(2011)}]{Eichmann:2011vu}
\bibinfo{author}{G.~Eichmann}, \bibinfo{title}{{Nucleon electromagnetic form
  factors from the covariant Faddeev equation}}, \bibinfo{journal}{Phys. Rev.
  D} \bibinfo{volume}{84} (\bibinfo{year}{2011}) \bibinfo{pages}{014014}.

\bibitem[{Yao et~al.(2025)Yao, Binosi, Cui, and Roberts}]{Yao:2024uej}
\bibinfo{author}{Z.-Q. Yao}, \bibinfo{author}{D.~Binosi},
  \bibinfo{author}{Z.-F. Cui}, \bibinfo{author}{C.~D. Roberts},
  \bibinfo{title}{{Nucleon charge and magnetisation distributions: flavour
  separation and zeroes -- arXiv:2403.08088 [hep-ph]}}, \bibinfo{journal}{Fund.
  Res.}  (\bibinfo{year}{2025}) \bibinfo{pages}{\emph{in
  press}}\bibinfo{note}{{
  \href{https://doi.org/10.1016/j.fmre.2024.11.005}{10.1016/j.fmre.2024.11.005}}}.

\bibitem[{Barabanov et~al.(2021)}]{Barabanov:2020jvn}
\bibinfo{author}{M.~Y. Barabanov}, et~al., \bibinfo{title}{{Diquark
  Correlations in Hadron Physics: Origin, Impact and Evidence}},
  \bibinfo{journal}{Prog. Part. Nucl. Phys.} \bibinfo{volume}{116}
  (\bibinfo{year}{2021}) \bibinfo{pages}{103835}.

\bibitem[{Cheng et~al.(2025)Cheng, Yao, Binosi, Lu, and
  Roberts}]{Cheng:2025yij}
\bibinfo{author}{P.~Cheng}, \bibinfo{author}{Z.~Q. Yao},
  \bibinfo{author}{D.~Binosi}, \bibinfo{author}{Y.~Lu}, \bibinfo{author}{C.~D.
  Roberts}, \bibinfo{title}{{Quark + diquark description of nucleon elastic
  electromagnetic form factors}}, \bibinfo{journal}{Eur. Phys. J. A}
  \bibinfo{volume}{61}~(\bibinfo{number}{11}) (\bibinfo{year}{2025})
  \bibinfo{pages}{255}.

\bibitem[{Sachs(1962)}]{Sachs:1962zzc}
\bibinfo{author}{R.~Sachs}, \bibinfo{title}{{High-Energy Behavior of Nucleon
  Electromagnetic Form Factors}}, \bibinfo{journal}{Phys. Rev.}
  \bibinfo{volume}{126} (\bibinfo{year}{1962}) \bibinfo{pages}{2256--2260}.

\bibitem[{Miller(2010)}]{Miller:2010nz}
\bibinfo{author}{G.~A. Miller}, \bibinfo{title}{{Transverse Charge Densities}},
  \bibinfo{journal}{Ann. Rev. Nucl. Part. Sci.} \bibinfo{volume}{60}
  (\bibinfo{year}{2010}) \bibinfo{pages}{1--25}.

\bibitem[{Drell and Yan(1970)}]{Drell:1969km}
\bibinfo{author}{S.~D. Drell}, \bibinfo{author}{T.-M. Yan},
  \bibinfo{title}{{Connection of Elastic Electromagnetic Nucleon Form-Factors
  at Large Q**2 and Deep Inelastic Structure Functions Near Threshold}},
  \bibinfo{journal}{Phys. Rev. Lett.} \bibinfo{volume}{24}
  (\bibinfo{year}{1970}) \bibinfo{pages}{181--185}.

\bibitem[{West(1970)}]{West:1970av}
\bibinfo{author}{G.~B. West}, \bibinfo{title}{{Phenomenological model for the
  electromagnetic structure of the proton}}, \bibinfo{journal}{Phys. Rev.
  Lett.} \bibinfo{volume}{24} (\bibinfo{year}{1970})
  \bibinfo{pages}{1206--1209}.

\bibitem[{Munczek(1995)}]{Munczek:1994zz}
\bibinfo{author}{H.~J. Munczek}, \bibinfo{title}{{Dynamical chiral symmetry
  breaking, Goldstone's theorem and the consistency of the Schwinger-Dyson and
  Bethe-Salpeter Equations}}, \bibinfo{journal}{Phys. Rev. D}
  \bibinfo{volume}{52} (\bibinfo{year}{1995}) \bibinfo{pages}{4736--4740}.

\bibitem[{Bender et~al.(1996)Bender, Roberts, and von Smekal}]{Bender:1996bb}
\bibinfo{author}{A.~Bender}, \bibinfo{author}{C.~D. Roberts},
  \bibinfo{author}{L.~von Smekal}, \bibinfo{title}{{Goldstone Theorem and
  Diquark Confinement Beyond Rainbow- Ladder Approximation}},
  \bibinfo{journal}{Phys. Lett. B} \bibinfo{volume}{380} (\bibinfo{year}{1996})
  \bibinfo{pages}{7--12}.

\bibitem[{Roberts et~al.(2021)Roberts, Richards, Horn, and
  Chang}]{Roberts:2021nhw}
\bibinfo{author}{C.~D. Roberts}, \bibinfo{author}{D.~G. Richards},
  \bibinfo{author}{T.~Horn}, \bibinfo{author}{L.~Chang},
  \bibinfo{title}{{Insights into the emergence of mass from studies of pion and
  kaon structure}}, \bibinfo{journal}{Prog. Part. Nucl. Phys.}
  \bibinfo{volume}{120} (\bibinfo{year}{2021}) \bibinfo{pages}{103883}.

\bibitem[{Raya et~al.(2024)Raya, Bashir, Binosi, Roberts, and
  Rodr\'\i{}guez-Quintero}]{Raya:2024ejx}
\bibinfo{author}{K.~Raya}, \bibinfo{author}{A.~Bashir},
  \bibinfo{author}{D.~Binosi}, \bibinfo{author}{C.~D. Roberts},
  \bibinfo{author}{J.~Rodr\'\i{}guez-Quintero}, \bibinfo{title}{{Pseudoscalar
  Mesons and Emergent Mass}}, \bibinfo{journal}{Few Body Syst.}
  \bibinfo{volume}{65}~(\bibinfo{number}{2}) (\bibinfo{year}{2024})
  \bibinfo{pages}{60}.

\bibitem[{Fischer and Williams(2009)}]{Fischer:2009jm}
\bibinfo{author}{C.~S. Fischer}, \bibinfo{author}{R.~Williams},
  \bibinfo{title}{{Probing the gluon self-interaction in light mesons}},
  \bibinfo{journal}{Phys. Rev. Lett.} \bibinfo{volume}{103}
  (\bibinfo{year}{2009}) \bibinfo{pages}{122001}.

\bibitem[{Chang and Roberts(2009)}]{Chang:2009zb}
\bibinfo{author}{L.~Chang}, \bibinfo{author}{C.~D. Roberts},
  \bibinfo{title}{{Sketching the Bethe-Salpeter kernel}},
  \bibinfo{journal}{Phys. Rev. Lett.} \bibinfo{volume}{103}
  (\bibinfo{year}{2009}) \bibinfo{pages}{081601}.

\bibitem[{Chang et~al.(2013)Chang, Cloet, Cobos-Martinez, Roberts, Schmidt, and
  Tandy}]{Chang:2013pq}
\bibinfo{author}{L.~Chang}, \bibinfo{author}{I.~C. Cloet},
  \bibinfo{author}{J.~J. Cobos-Martinez}, \bibinfo{author}{C.~D. Roberts},
  \bibinfo{author}{S.~M. Schmidt}, \bibinfo{author}{P.~C. Tandy},
  \bibinfo{title}{{Imaging dynamical chiral symmetry breaking: pion wave
  function on the light front}}, \bibinfo{journal}{Phys. Rev. Lett.}
  \bibinfo{volume}{110} (\bibinfo{year}{2013}) \bibinfo{pages}{132001}.

\bibitem[{Binosi et~al.(2015)Binosi, Chang, Papavassiliou, and
  Roberts}]{Binosi:2014aea}
\bibinfo{author}{D.~Binosi}, \bibinfo{author}{L.~Chang},
  \bibinfo{author}{J.~Papavassiliou}, \bibinfo{author}{C.~D. Roberts},
  \bibinfo{title}{{Bridging a gap between continuum-QCD and \emph{ab initio}
  predictions of hadron observables}}, \bibinfo{journal}{Phys. Lett. B}
  \bibinfo{volume}{742} (\bibinfo{year}{2015}) \bibinfo{pages}{183--188}.

\bibitem[{Williams et~al.(2016)Williams, Fischer, and
  Heupel}]{Williams:2015cvx}
\bibinfo{author}{R.~Williams}, \bibinfo{author}{C.~S. Fischer},
  \bibinfo{author}{W.~Heupel}, \bibinfo{title}{{Light mesons in QCD and
  unquenching effects from the 3PI effective action}}, \bibinfo{journal}{Phys.
  Rev. D} \bibinfo{volume}{93} (\bibinfo{year}{2016}) \bibinfo{pages}{034026}.

\bibitem[{Binosi et~al.(2016)Binosi, Chang, Qin, Papavassiliou, and
  Roberts}]{Binosi:2016rxz}
\bibinfo{author}{D.~Binosi}, \bibinfo{author}{L.~Chang}, \bibinfo{author}{S.-X.
  Qin}, \bibinfo{author}{J.~Papavassiliou}, \bibinfo{author}{C.~D. Roberts},
  \bibinfo{title}{{Symmetry preserving truncations of the gap and
  Bethe-Salpeter equations}}, \bibinfo{journal}{Phys. Rev. D}
  \bibinfo{volume}{93} (\bibinfo{year}{2016}) \bibinfo{pages}{096010}.

\bibitem[{Binosi et~al.(2017)Binosi, Chang, Papavassiliou, Qin, and
  Roberts}]{Binosi:2016wcx}
\bibinfo{author}{D.~Binosi}, \bibinfo{author}{L.~Chang},
  \bibinfo{author}{J.~Papavassiliou}, \bibinfo{author}{S.-X. Qin},
  \bibinfo{author}{C.~D. Roberts}, \bibinfo{title}{{Natural constraints on the
  gluon-quark vertex}}, \bibinfo{journal}{Phys. Rev. D} \bibinfo{volume}{95}
  (\bibinfo{year}{2017}) \bibinfo{pages}{031501(R)}.

\bibitem[{Xu et~al.(2023)Xu, Yao, Qin, Cui, and Roberts}]{Xu:2022kng}
\bibinfo{author}{Z.-N. Xu}, \bibinfo{author}{Z.-Q. Yao}, \bibinfo{author}{S.-X.
  Qin}, \bibinfo{author}{Z.-F. Cui}, \bibinfo{author}{C.~D. Roberts},
  \bibinfo{title}{{Bethe-Salpeter kernel and properties of strange-quark
  mesons}}, \bibinfo{journal}{Eur. Phys. J. A}
  \bibinfo{volume}{59}~(\bibinfo{number}{3}) (\bibinfo{year}{2023})
  \bibinfo{pages}{39}.

\bibitem[{Chen et~al.(2022)Chen, Fischer, Roberts, and Segovia}]{Chen:2021guo}
\bibinfo{author}{C.~Chen}, \bibinfo{author}{C.~S. Fischer},
  \bibinfo{author}{C.~D. Roberts}, \bibinfo{author}{J.~Segovia},
  \bibinfo{title}{{Nucleon axial-vector and pseudoscalar form factors and PCAC
  relations}}, \bibinfo{journal}{Phys. Rev. D}
  \bibinfo{volume}{105}~(\bibinfo{number}{9}) (\bibinfo{year}{2022})
  \bibinfo{pages}{094022}.

\bibitem[{Chang and Roberts(2021)}]{Chang:2021utv}
\bibinfo{author}{L.~Chang}, \bibinfo{author}{C.~D. Roberts},
  \bibinfo{title}{{Regarding the distribution of glue in the pion}},
  \bibinfo{journal}{Chin. Phys. Lett.}
  \bibinfo{volume}{38}~(\bibinfo{number}{8}) (\bibinfo{year}{2021})
  \bibinfo{pages}{081101}.

\bibitem[{Lu et~al.(2024)Lu, Xu, Raya, Roberts, and
  Rodr\'\i{}guez-Quintero}]{Lu:2023yna}
\bibinfo{author}{Y.~Lu}, \bibinfo{author}{Y.-Z. Xu}, \bibinfo{author}{K.~Raya},
  \bibinfo{author}{C.~D. Roberts},
  \bibinfo{author}{J.~Rodr\'\i{}guez-Quintero}, \bibinfo{title}{{Pion
  distribution functions from low-order Mellin moments}},
  \bibinfo{journal}{Phys. Lett. B} \bibinfo{volume}{850} (\bibinfo{year}{2024})
  \bibinfo{pages}{138534}.

\bibitem[{Chen et~al.(2024)Chen, Fischer, and Roberts}]{Chen:2023zhh}
\bibinfo{author}{C.~Chen}, \bibinfo{author}{C.~S. Fischer},
  \bibinfo{author}{C.~D. Roberts}, \bibinfo{title}{{Nucleon to $\Delta$ axial
  and pseudoscalar transition form factors}}, \bibinfo{journal}{Phys. Rev.
  Lett.} \bibinfo{volume}{133}~(\bibinfo{number}{13}) (\bibinfo{year}{2024})
  \bibinfo{pages}{131901}.

\bibitem[{Yu et~al.(2024)Yu, Cheng, Xing, Gao, and Roberts}]{Yu:2024qsd}
\bibinfo{author}{Y.~Yu}, \bibinfo{author}{P.~Cheng}, \bibinfo{author}{H.-Y.
  Xing}, \bibinfo{author}{F.~Gao}, \bibinfo{author}{C.~D. Roberts},
  \bibinfo{title}{{Contact interaction study of proton parton distributions}},
  \bibinfo{journal}{Eur. Phys. J. C} \bibinfo{volume}{84}~(\bibinfo{number}{7})
  (\bibinfo{year}{2024}) \bibinfo{pages}{739}.

\bibitem[{Alexandrou et~al.(2025)}]{Alexandrou:2024zvn}
\bibinfo{author}{C.~Alexandrou}, et~al., \bibinfo{title}{{Quark and Gluon
  Momentum Fractions in the Pion and in the Kaon}}, \bibinfo{journal}{Phys.
  Rev. Lett.} \bibinfo{volume}{134}~(\bibinfo{number}{13})
  (\bibinfo{year}{2025}) \bibinfo{pages}{131902}.

\bibitem[{Maris and Roberts(1997)}]{Maris:1997tm}
\bibinfo{author}{P.~Maris}, \bibinfo{author}{C.~D. Roberts},
  \bibinfo{title}{{{$\pi$} and {$K$} meson Bethe-Salpeter amplitudes}},
  \bibinfo{journal}{Phys. Rev. C} \bibinfo{volume}{56} (\bibinfo{year}{1997})
  \bibinfo{pages}{3369--3383}.

\bibitem[{Bashir et~al.(2009)Bashir, Raya, S{\'a}nchez-Madrigal, and
  Roberts}]{Bashir:2009fv}
\bibinfo{author}{A.~Bashir}, \bibinfo{author}{A.~Raya},
  \bibinfo{author}{S.~S{\'a}nchez-Madrigal}, \bibinfo{author}{C.~D. Roberts},
  \bibinfo{title}{{Gauge invariance of a critical number of flavours in QED3}},
  \bibinfo{journal}{Few Body Syst.} \bibinfo{volume}{46} (\bibinfo{year}{2009})
  \bibinfo{pages}{229--237}.

\bibitem[{Qin et~al.(2011)Qin, Chang, Liu, Roberts, and Wilson}]{Qin:2011dd}
\bibinfo{author}{S.-X. Qin}, \bibinfo{author}{L.~Chang}, \bibinfo{author}{Y.-X.
  Liu}, \bibinfo{author}{C.~D. Roberts}, \bibinfo{author}{D.~J. Wilson},
  \bibinfo{title}{{Interaction model for the gap equation}},
  \bibinfo{journal}{Phys. Rev. C} \bibinfo{volume}{84} (\bibinfo{year}{2011})
  \bibinfo{pages}{042202(R)}.

\bibitem[{Chang et~al.(2009)Chang, Liu, Roberts, Shi, Sun, and
  Zong}]{Chang:2008ec}
\bibinfo{author}{L.~Chang}, \bibinfo{author}{Y.-X. Liu}, \bibinfo{author}{C.~D.
  Roberts}, \bibinfo{author}{Y.-M. Shi}, \bibinfo{author}{W.-M. Sun},
  \bibinfo{author}{H.-S. Zong}, \bibinfo{title}{{Chiral susceptibility and the
  scalar Ward identity}}, \bibinfo{journal}{Phys. Rev. C} \bibinfo{volume}{79}
  (\bibinfo{year}{2009}) \bibinfo{pages}{035209}.

\bibitem[{Navas et~al.(2024)}]{ParticleDataGroup:2024cfk}
\bibinfo{author}{S.~Navas}, et~al., \bibinfo{title}{{Review of particle
  physics}}, \bibinfo{journal}{Phys. Rev. D}
  \bibinfo{volume}{110}~(\bibinfo{number}{3}) (\bibinfo{year}{2024})
  \bibinfo{pages}{030001}.

\bibitem[{Qin and Roberts(2020)}]{Qin:2020rad}
\bibinfo{author}{S.-X. Qin}, \bibinfo{author}{C.~D. Roberts},
  \bibinfo{title}{{Impressions of the Continuum Bound State Problem in QCD}},
  \bibinfo{journal}{Chin. Phys. Lett.}
  \bibinfo{volume}{37}~(\bibinfo{number}{12}) (\bibinfo{year}{2020})
  \bibinfo{pages}{121201}.

\bibitem[{Cui et~al.(2020)Cui, Zhang, Binosi, de~Soto, Mezrag, Papavassiliou,
  Roberts, Rodr{\'{\i}}guez-Quintero, Segovia, and Zafeiropoulos}]{Cui:2019dwv}
\bibinfo{author}{Z.-F. Cui}, \bibinfo{author}{J.-L. Zhang},
  \bibinfo{author}{D.~Binosi}, \bibinfo{author}{F.~de~Soto},
  \bibinfo{author}{C.~Mezrag}, \bibinfo{author}{J.~Papavassiliou},
  \bibinfo{author}{C.~D. Roberts},
  \bibinfo{author}{J.~Rodr{\'{\i}}guez-Quintero}, \bibinfo{author}{J.~Segovia},
  \bibinfo{author}{S.~Zafeiropoulos}, \bibinfo{title}{{Effective charge from
  lattice QCD}}, \bibinfo{journal}{Chin. Phys. C} \bibinfo{volume}{44}
  (\bibinfo{year}{2020}) \bibinfo{pages}{083102}.

\bibitem[{Deur et~al.(2024)Deur, Brodsky, and Roberts}]{Deur:2023dzc}
\bibinfo{author}{A.~Deur}, \bibinfo{author}{S.~J. Brodsky},
  \bibinfo{author}{C.~D. Roberts}, \bibinfo{title}{{QCD Running Couplings and
  Effective Charges}}, \bibinfo{journal}{Prog. Part. Nucl. Phys.}
  \bibinfo{volume}{134} (\bibinfo{year}{2024}) \bibinfo{pages}{104081}.

\bibitem[{Brodsky et~al.(2024)Brodsky, Deur, and Roberts}]{Brodsky:2024zev}
\bibinfo{author}{S.~J. Brodsky}, \bibinfo{author}{A.~Deur},
  \bibinfo{author}{C.~D. Roberts}, \bibinfo{title}{{The Secret to the Strongest
  Force in the Universe}}, \bibinfo{journal}{Sci. Am.} \bibinfo{volume}{5
  (May)} (\bibinfo{year}{2024}) \bibinfo{pages}{32--39}.

\bibitem[{Maris and Tandy(2006)}]{Maris:2005tt}
\bibinfo{author}{P.~Maris}, \bibinfo{author}{P.~C. Tandy}, \bibinfo{title}{{QCD
  modeling of hadron physics}}, \bibinfo{journal}{Nucl. Phys. Proc. Suppl.}
  \bibinfo{volume}{161} (\bibinfo{year}{2006}) \bibinfo{pages}{136--152}.

\bibitem[{Krassnigg(2008)}]{Krassnigg:2009gd}
\bibinfo{author}{A.~Krassnigg}, \bibinfo{title}{{Excited mesons in a
  Bethe-Salpeter approach}}, \bibinfo{journal}{PoS}
  \bibinfo{volume}{CONFINEMENT\,8} (\bibinfo{year}{2008}) \bibinfo{pages}{075}.

\bibitem[{Eichmann and Fischer(2012)}]{Eichmann:2011pv}
\bibinfo{author}{G.~Eichmann}, \bibinfo{author}{C.~S. Fischer},
  \bibinfo{title}{{Nucleon axial and pseudoscalar form factors from the
  covariant Faddeev equation}}, \bibinfo{journal}{Eur. Phys. J. A}
  \bibinfo{volume}{48} (\bibinfo{year}{2012}) \bibinfo{pages}{9}.

\bibitem[{Xu et~al.(2019)Xu, Binosi, Cui, Li, Roberts, Xu, and
  Zong}]{Xu:2019ilh}
\bibinfo{author}{Y.-Z. Xu}, \bibinfo{author}{D.~Binosi}, \bibinfo{author}{Z.-F.
  Cui}, \bibinfo{author}{B.-L. Li}, \bibinfo{author}{C.~D. Roberts},
  \bibinfo{author}{S.-S. Xu}, \bibinfo{author}{H.-S. Zong},
  \bibinfo{title}{{Elastic electromagnetic form factors of vector mesons}},
  \bibinfo{journal}{Phys. Rev. D} \bibinfo{volume}{100} (\bibinfo{year}{2019})
  \bibinfo{pages}{114038}.

\bibitem[{Eichmann et~al.(2010)Eichmann, Alkofer, Krassnigg, and
  Nicmorus}]{Eichmann:2009en}
\bibinfo{author}{G.~Eichmann}, \bibinfo{author}{R.~Alkofer},
  \bibinfo{author}{A.~Krassnigg}, \bibinfo{author}{D.~Nicmorus},
  \bibinfo{title}{{Covariant solution of the three-quark problem in quantum
  field theory: The Nucleon}}, \bibinfo{journal}{EPJ Web Conf.}
  \bibinfo{volume}{3} (\bibinfo{year}{2010}) \bibinfo{pages}{03028}.

\bibitem[{Ye et~al.(2018)Ye, Arrington, Hill, and Lee}]{Ye:2017gyb}
\bibinfo{author}{Z.~Ye}, \bibinfo{author}{J.~Arrington}, \bibinfo{author}{R.~J.
  Hill}, \bibinfo{author}{G.~Lee}, \bibinfo{title}{{Proton and Neutron
  Electromagnetic Form Factors and Uncertainties}}, \bibinfo{journal}{Phys.
  Lett. B} \bibinfo{volume}{777} (\bibinfo{year}{2018}) \bibinfo{pages}{8--15}.

\bibitem[{Kelly(2004)}]{Kelly:2004hm}
\bibinfo{author}{J.~J. Kelly}, \bibinfo{title}{{Simple parametrization of
  nucleon form factors}}, \bibinfo{journal}{Phys. Rev. C} \bibinfo{volume}{70}
  (\bibinfo{year}{2004}) \bibinfo{pages}{068202}.

\bibitem[{Xu et~al.(2021{\natexlab{b}})Xu, Chen, Yao, Binosi, Cui, and
  Roberts}]{Xu:2021mju}
\bibinfo{author}{Y.-Z. Xu}, \bibinfo{author}{S.~Chen}, \bibinfo{author}{Z.-Q.
  Yao}, \bibinfo{author}{D.~Binosi}, \bibinfo{author}{Z.-F. Cui},
  \bibinfo{author}{C.~D. Roberts}, \bibinfo{title}{{Vector-meson production and
  vector meson dominance}}, \bibinfo{journal}{Eur. Phys. J. C}
  \bibinfo{volume}{81} (\bibinfo{year}{2021}{\natexlab{b}})
  \bibinfo{pages}{895}.

\bibitem[{Wilson et~al.(2012)Wilson, Cloet, Chang, and Roberts}]{Wilson:2011aa}
\bibinfo{author}{D.~J. Wilson}, \bibinfo{author}{I.~C. Cloet},
  \bibinfo{author}{L.~Chang}, \bibinfo{author}{C.~D. Roberts},
  \bibinfo{title}{{Nucleon and Roper electromagnetic elastic and transition
  form factors}}, \bibinfo{journal}{Phys. Rev. C} \bibinfo{volume}{85}
  (\bibinfo{year}{2012}) \bibinfo{pages}{025205}.

\bibitem[{Yin et~al.(2023)Yin, Xu, Cui, Roberts, and
  Rodr\'\i{}guez-Quintero}]{Yin:2023dbw}
\bibinfo{author}{P.-L. Yin}, \bibinfo{author}{Y.-Z. Xu}, \bibinfo{author}{Z.-F.
  Cui}, \bibinfo{author}{C.~D. Roberts},
  \bibinfo{author}{J.~Rodr\'\i{}guez-Quintero}, \bibinfo{title}{{All-Orders
  Evolution of Parton Distributions: Principle, Practice, and Predictions}},
  \bibinfo{journal}{Chin. Phys. Lett. \emph{Express}}
  \bibinfo{volume}{40}~(\bibinfo{number}{9}) (\bibinfo{year}{2023})
  \bibinfo{pages}{091201}.

\bibitem[{Close and Thomas(1988)}]{Close:1988br}
\bibinfo{author}{F.~E. Close}, \bibinfo{author}{A.~W. Thomas},
  \bibinfo{title}{{The Spin and Flavor Dependence of Parton Distribution
  Functions}}, \bibinfo{journal}{Phys. Lett. B} \bibinfo{volume}{212}
  (\bibinfo{year}{1988}) \bibinfo{pages}{227}.

\bibitem[{Cui et~al.(2022{\natexlab{b}})Cui, Gao, Binosi, Chang, Roberts, and
  Schmidt}]{Cui:2021gzg}
\bibinfo{author}{Z.-F. Cui}, \bibinfo{author}{F.~Gao},
  \bibinfo{author}{D.~Binosi}, \bibinfo{author}{L.~Chang},
  \bibinfo{author}{C.~D. Roberts}, \bibinfo{author}{S.~M. Schmidt},
  \bibinfo{title}{{Valence quark ratio in the proton}}, \bibinfo{journal}{Chin.
  Phys. Lett. \emph{Express}} \bibinfo{volume}{39}~(\bibinfo{number}{04})
  (\bibinfo{year}{2022}{\natexlab{b}}) \bibinfo{pages}{041401}.

\bibitem[{Cheng et~al.(2023)Cheng, Yu, Xing, Chen, Cui, and
  Roberts}]{Cheng:2023kmt}
\bibinfo{author}{P.~Cheng}, \bibinfo{author}{Y.~Yu}, \bibinfo{author}{H.-Y.
  Xing}, \bibinfo{author}{C.~Chen}, \bibinfo{author}{Z.-F. Cui},
  \bibinfo{author}{C.~D. Roberts}, \bibinfo{title}{{Perspective on polarised
  parton distribution functions and proton spin}}, \bibinfo{journal}{Phys.
  Lett. B} \bibinfo{volume}{844} (\bibinfo{year}{2023})
  \bibinfo{pages}{138074}.

\bibitem[{Schlessinger and Schwartz(1966)}]{Schlessinger:1966zz}
\bibinfo{author}{L.~Schlessinger}, \bibinfo{author}{C.~Schwartz},
  \bibinfo{title}{{Analyticity as a Useful Computation Tool}},
  \bibinfo{journal}{Phys. Rev. Lett.} \bibinfo{volume}{16}
  (\bibinfo{year}{1966}) \bibinfo{pages}{1173--1174}.

\bibitem[{Schlessinger(1968)}]{PhysRev.167.1411}
\bibinfo{author}{L.~Schlessinger}, \bibinfo{title}{Use of Analyticity in the
  Calculation of Nonrelativistic Scattering Amplitudes},
  \bibinfo{journal}{Phys. Rev.} \bibinfo{volume}{167} (\bibinfo{year}{1968})
  \bibinfo{pages}{1411--1423}.

\bibitem[{Tripolt et~al.(2017)Tripolt, Haritan, Wambach, and
  Moiseyev}]{Tripolt:2016cya}
\bibinfo{author}{R.~A. Tripolt}, \bibinfo{author}{I.~Haritan},
  \bibinfo{author}{J.~Wambach}, \bibinfo{author}{N.~Moiseyev},
  \bibinfo{title}{{Threshold energies and poles for hadron physical problems by
  a model-independent universal algorithm}}, \bibinfo{journal}{Phys. Lett. B}
  \bibinfo{volume}{774} (\bibinfo{year}{2017}) \bibinfo{pages}{411--416}.

\bibitem[{Cahill et~al.(1989)Cahill, Roberts, and Praschifka}]{Cahill:1988dx}
\bibinfo{author}{R.~T. Cahill}, \bibinfo{author}{C.~D. Roberts},
  \bibinfo{author}{J.~Praschifka}, \bibinfo{title}{{Baryon structure and QCD}},
  \bibinfo{journal}{Austral. J. Phys.} \bibinfo{volume}{42}
  (\bibinfo{year}{1989}) \bibinfo{pages}{129--145}.

\bibitem[{Reinhardt(1990)}]{Reinhardt:1989rw}
\bibinfo{author}{H.~Reinhardt}, \bibinfo{title}{{Hadronization of Quark Flavor
  Dynamics}}, \bibinfo{journal}{Phys. Lett. B} \bibinfo{volume}{244}
  (\bibinfo{year}{1990}) \bibinfo{pages}{316--326}.

\bibitem[{Efimov et~al.(1990)Efimov, Ivanov, and Lyubovitskij}]{Efimov:1990uz}
\bibinfo{author}{G.~V. Efimov}, \bibinfo{author}{M.~A. Ivanov},
  \bibinfo{author}{V.~E. Lyubovitskij}, \bibinfo{title}{{Quark - diquark
  approximation of the three quark structure of baryons in the quark
  confinement model}}, \bibinfo{journal}{Z. Phys. C} \bibinfo{volume}{47}
  (\bibinfo{year}{1990}) \bibinfo{pages}{583--594}.

\bibitem[{Roberts(2020)}]{Roberts:2020hiw}
\bibinfo{author}{C.~D. Roberts}, \bibinfo{title}{{Empirical Consequences of
  Emergent Mass}}, \bibinfo{journal}{Symmetry} \bibinfo{volume}{12}
  (\bibinfo{year}{2020}) \bibinfo{pages}{1468}.

\bibitem[{Yu and Roberts(2024)}]{Yu:2024ovn}
\bibinfo{author}{Y.~Yu}, \bibinfo{author}{C.~D. Roberts},
  \bibinfo{title}{{Impressions of Parton Distribution Functions}},
  \bibinfo{journal}{Chin. Phys. Lett.} \bibinfo{volume}{41}
  (\bibinfo{year}{2024}) \bibinfo{pages}{121202}.

\bibitem[{Eichmann et~al.(2025)Eichmann, Fischer, and
  Hoffer}]{Eichmann:2025tzm}
\bibinfo{author}{G.~Eichmann}, \bibinfo{author}{C.~S. Fischer},
  \bibinfo{author}{J.~Hoffer}, \bibinfo{title}{{Functional methods for hadron
  spectroscopy\,--\,arXiv:2503.20718 [hep-ph]}}, in: \bibinfo{booktitle}{{16th
  Conference on Quark Confinement and the Hadron Spectrum}},
  \bibinfo{year}{2025}.

\bibitem[{Yu et~al.(2025)Yu, Cheng, Xing, Binosi, and Roberts}]{Yu:2025fer}
\bibinfo{author}{Y.~Yu}, \bibinfo{author}{P.~Cheng}, \bibinfo{author}{H.-Y.
  Xing}, \bibinfo{author}{D.~Binosi}, \bibinfo{author}{C.~D. Roberts},
  \bibinfo{title}{{Distribution Functions of $\Lambda$ and $\Sigma^0$
  Baryons}}, \bibinfo{journal}{Eur. Phys. J. A}
  \bibinfo{volume}{61}~(\bibinfo{number}{9}) (\bibinfo{year}{2025})
  \bibinfo{pages}{208}.

\bibitem[{Miller(2007)}]{Miller:2007uy}
\bibinfo{author}{G.~A. Miller}, \bibinfo{title}{{Charge Density of the
  Neutron}}, \bibinfo{journal}{Phys. Rev. Lett.} \bibinfo{volume}{99}
  (\bibinfo{year}{2007}) \bibinfo{pages}{112001}.

\bibitem[{Chang et~al.(2011)Chang, Liu, and Roberts}]{Chang:2010hb}
\bibinfo{author}{L.~Chang}, \bibinfo{author}{Y.-X. Liu}, \bibinfo{author}{C.~D.
  Roberts}, \bibinfo{title}{{Dressed-quark anomalous magnetic moments}},
  \bibinfo{journal}{Phys. Rev. Lett.} \bibinfo{volume}{106}
  (\bibinfo{year}{2011}) \bibinfo{pages}{072001}.

\bibitem[{Burkardt(2003)}]{Burkardt:2002hr}
\bibinfo{author}{M.~Burkardt}, \bibinfo{title}{{Impact parameter space
  interpretation for generalized parton distributions}}, \bibinfo{journal}{Int.
  J. Mod. Phys. A} \bibinfo{volume}{18} (\bibinfo{year}{2003})
  \bibinfo{pages}{173--208}.

\bibitem[{Carlson and Vanderhaeghen(2008)}]{Carlson:2007xd}
\bibinfo{author}{C.~E. Carlson}, \bibinfo{author}{M.~Vanderhaeghen},
  \bibinfo{title}{{Empirical transverse charge densities in the nucleon and the
  nucleon-to-Delta transition}}, \bibinfo{journal}{Phys. Rev. Lett.}
  \bibinfo{volume}{100} (\bibinfo{year}{2008}) \bibinfo{pages}{032004}.

\end{thebibliography}

\end{document}